\documentclass[12pt]{article}

\RequirePackage{geometry}
\usepackage{amsmath}
\usepackage[dvipsnames]{xcolor}
\usepackage{graphicx,psfrag,epsf}
\usepackage{enumerate}
\usepackage{natbib}
\usepackage[lined,ruled,vlined]{algorithm2e}
\usepackage{url} 
\usepackage{mathtools,amssymb,amsthm}
\usepackage{soul} 
\usepackage{float} 
\usepackage[colorlinks=false]{hyperref}

\newcommand{\Rhat}{{\widehat R}}

\newcommand{\ytilde}{{\widetilde y}}

\newcommand{\bw}{\mathbf{ w}}
\newcommand{\bx}{\mathbf{ x}}

\newcommand{\bz}{\mathbf{ z}}

\newcommand{\bV}{\mathbf{ V}}

\newcommand{\bX}{\mathbf{ X}}

\newcommand{\bZ}{\mathbf{ Z}}
\newcommand{\Dsc}{{\mathcal D}}

\newcommand{\Lsc}{{\mathcal L}}

\newcommand{\bbI}{\mathbb{ I}}

\newcommand{\balpha}{\boldsymbol{\alpha}}
\newcommand{\bbeta}{\boldsymbol{\beta}}

\newcommand{\btau}{\boldsymbol{\tau}}

\newcommand{\bphi}{\boldsymbol{\phi}}

\newcommand{\bpsi}{\boldsymbol{\psi}}

\newcommand{\bzero}{{\mathbf 0}}

\newcommand{\transpose}{{\mathsf{\scriptscriptstyle T}}}
\newcommand{\trans}{^{\transpose}}

\newcommand{\Dirichlet}{\text{Dirichlet}}

\newcommand{\MVN}{\text{MVN}}
\mathtoolsset{showonlyrefs=true} 
\def\spacingset#1{\renewcommand{\baselinestretch}%
{#1}\small\normalsize} \spacingset{1}

\theoremstyle{plain}

\theoremstyle{definition}

\begin{document}

\def\mytitle{Characterizing Quantile-varying Covariate Effects under the Accelerated Failure Time Model}
\title{\bf \mytitle}
  \author{Harrison T. Reeder\hspace{.2cm}\\
  Massachusetts General Hospital Biostatistics \\ 
  Department of Medicine, Harvard Medical School \\
  Kyu Ha Lee  \\ 
  Departments of Nutrition, Biostatistics, and Epidemiology, \\
  Harvard T.H. Chan School of Public Health \\
  Sebastien Haneuse\\
    Department of Biostatistics, Harvard T.H. Chan School of Public Health}
  \date{}
  \maketitle

\begin{abstract}
\noindent An important task in survival analysis is choosing a structure for the relationship between covariates of interest and the time-to-event outcome. For example, the accelerated failure time (AFT) model structures each covariate effect as a constant multiplicative shift in the outcome distribution across all survival quantiles. Though parsimonious, this structure cannot detect or capture effects that differ across quantiles of the distribution, a limitation that is analogous to only permitting proportional hazards in the Cox model. To address this, we propose a general framework for quantile-varying multiplicative effects under the AFT model. Specifically, we embed flexible regression structures within the AFT model, and derive a novel formula for interpretable effects on the quantile scale. A regression standardization scheme based on the g-formula is proposed to enable estimation of both covariate-conditional and marginal effects for an exposure of interest. We implement a user-friendly Bayesian approach for estimation and quantification of uncertainty, while accounting for left truncation and complex censoring. We emphasize the intuitive interpretation of this model through numerical and graphical tools, and illustrate its performance through simulation and application to a study of Alzheimer's disease and dementia.
\end{abstract}

\noindent%
{\it Keywords: Accelerated failure time model; Bayesian survival analysis; Left-truncation;  Time-varying coefficients; Time-varying covariates}

\vspace{0.3in}
{\tiny
\noindent This is the final "accepted" version of the article is published in \textit{Biostatistics} by Oxford University Press. It has been uploaded following an embargo period from original publication \\

\noindent Reeder HT, Lee KH, Haneuse S. Characterizing quantile-varying covariate effects under the accelerated failure time model. Biostatistics. 2024 Apr 15;25(2):449-467. doi: \href{https://doi.org/10.1093/biostatistics/kxac052}{10.1093/biostatistics/kxac052}. PMID: \href{https://pubmed.ncbi.nlm.nih.gov/36610077/}{36610077}; PMCID: PMC11484523.


\newpage
\spacingset{1.15} 


\section{Introduction}
\label{sec:intro}

Modeling the relationship between a time-to-event outcome $T$ and a vector of covariates $\bX$ requires choosing a structure for the covariate effects. The proportional hazards model is by far the most commonly used model, specifying a constant multiplicative effect on the hazard of the outcome, yielding a `hazard ratio.' Though ubiquitous, hazard ratios can be difficult to interpret, and the constant effect across time---that is, the `proportionality' of the hazards---is not always plausible \citep{hernan2010hazards,uno2015alternatives}. As an alternative, the accelerated failure time (AFT) model directly describes shifts in the outcome distribution between populations having different characteristics, via multiplicative effects on event time quantiles \citep{wei1992accelerated}. Specifically, every survival quantile is multiplied by a constant `acceleration factor,' equivalent to a horizontal stretching or compressing of the survivor function. In other words, the times by which 10 percent of events occur, or 50 percent (i.e., the median survival time), or any other quantile, are shifted by the same multiplicative constant. This common effect across quantiles is the central feature of the AFT model, making it highly interpretable because contrasts of survival quantiles are tangible and often clinically meaningful. Despite the parsimony of a constant multiplicative effect, in some settings it may be important to allow for more flexible effects across quantiles. For example, consider the study of Alzheimer's disease (AD) and dementia among older adults. Prospective cohort studies of incident AD and dementia typically enroll subjects and follow them over decades, often subject to left truncation and sometimes complex censoring. Age at AD onset among those with a particular risk factor, for example, may skew earlier than among those without the risk factor. However, because AD is a complex disease that can arise over a long time scale, baseline risk factors may not affect the entire distribution uniformly. For example, a risk factor may not affect the timing of `early-onset' cases, but make `late-onset' cases occur sooner.

Modeling flexibly time-varying hazard ratios is a well-known and commonly used tool under the Cox model, but analogous extensions to flexibly quantile-varying acceleration factors under the AFT model are less well-studied. Some extensions to the AFT include admitting covariates to accommodate `heteroskedasticity' or stratification of the baseline distribution \citep{hsieh2001heteroscedastic,zhou2017generalized}. Recent work by \citet{crowther2022flexible} describes a frequentist spline-based AFT model, with possibly time-varying covariate effects. However, their work considers a less common interpretation of the acceleration factor on the scale of log time, rather than investigating the potential for flexibility on the quantile scale. Separately, \citet{pang2021flexible} consider a frequentist spline-based AFT model using a completely different formulation derived from \citet{prentice1979hazard}, requiring a specialized estimation algorithm and bootstrapping for inference. Neither of these these papers explore the Bayesian paradigm or examine effects of time-dependent covariates that commonly arise in longitudinal studies, for which the resulting relationship varies both over the trajectory of the covariate, and over the survival quantiles.

In this paper, we extend the AFT model to allow flexible acceleration factors that vary across quantiles. Our approach builds on a time-varying AFT model first introduced in \citet{cox1984analysis} but seemingly largely overlooked in the literature, and a general framework for flexible covariate effect specification. AFT regression coefficients specified to vary over time can be inverted into quantile-varying acceleration factors, and we develop a regression standardization scheme based on the g-formula to estimate marginal acceleration factors for an exposure. We propose a Bayesian estimation approach using the Stan language, which allows quantification of uncertainty and increased flexibility. Through simulations, we examine the ability for the proposed flexible AFT effect specification and regression standardization methods to recover true effects in multiple settings, and illustrate potential sensitivity to baseline hazard specification.

Finally, we derive new insights into the use of binary time-varying covariates under the AFT model, and present novel tools for modeling and visualizing such effects. This further expands the AFT modeling toolkit to cover many extensions commonly used under the Cox model. We motivate these methods with an in-depth analysis of the Religious Orders Study and Memory and Aging Project prospective cohort studies of AD and dementia \citep{bennett2018religious}.

\section{The Accelerated Failure Time Model}

The standard AFT model with time-invariant effects can be written as a log-linear model of time:
\begin{align}
\log(T) & = \bX\trans\bbeta + \epsilon,
\end{align}
where $\epsilon$ is a random error term and $\bbeta$ is a vector of regression coefficients corresponding with covariates $\bX$.
We denote the exponentiated error $T_0 = \exp(\epsilon)$, representing a hypothetical random variable drawn from the ``baseline distribution'' having survivor function $S_0$.
This model structures covariate effects such that the distribution of event times among subjects having covariate pattern $\bx$, denoted $T_{\bx}$, is directly shifted from the baseline distribution by the transformation
\begin{align}
T_{\bx} \times \exp(-\bx\trans\bbeta)  \sim S_0. 
\end{align}
Based on this connection, an equivalent representation of the standard AFT model is given directly via the baseline survivor function $S_0$ as
\begin{align}\label{eq:aftsurv}
S(t\mid\bX) & = S_0(t \times \exp(-\bX\trans\bbeta)).
\end{align}

The AFT model admits interpretation of covariate effects as multiplicative shifts of survival quantiles. Defining $t_{\bx}^{(p)}$ and $t_{0}^{(p)}$ to be the $p$th quantile times under $\bx$ and baseline respectively,
\begin{align}
p & = S(t^{(p)}_{\bx}\mid\bx)  = S_0(t^{(p)}_{\bx} \times \exp(-\bx\trans\bbeta)) =  S_0(t^{(p)}_{0}).
\end{align}
Solving for the $p$th quantile survival time under $\bx$ yields
\begin{align}
t^{(p)}_{\bx} = S^{-1}(p \mid \bx) = S_0^{-1}(p)\exp(\bx\trans\bbeta).
\end{align}
The acceleration factor between two covariate patterns $\bx$ and $\bx'$ is then the ratio of $p$th quantiles,
\begin{align}
t^{(p)}_{\bx} / t^{(p)}_{\bx'} = \frac{ S_0^{-1}(p)\exp(\bx\trans\bbeta)}{ S_0^{-1}(p)\exp((\bx')\trans\bbeta)} =\exp((\bx-\bx')\trans\bbeta).
\end{align}
Under the standard AFT model, the acceleration factor does not depend on the form of $S_0$ or the value of $p$, characterizing a constant multiplicative covariate effect across the entire distribution.

\subsection{AFT model with time-varying components}
In the standard AFT model \eqref{eq:aftsurv}, the covariate-adjusted survivor function is characterized by the time shift $t \times \exp(-\bX\trans\bbeta)$. Towards a more flexible AFT model, we replace this time shift with a general increasing function $V(t\mid\bX)$, yielding the covariate-adjusted survivor function
\begin{align}\label{eq:aftsurv2}
S(t\mid\bX) & = S_0\left( V(t\mid\bX) \right).
\end{align}
This formulation, first discussed by \citet{cox1984analysis} in the context of time-varying covariates, reduces to the standard AFT when $V(t\mid\bX) = t\times\exp(-\bX\trans\bbeta)$, while also admitting other temporal specifications of the relationship between covariates and the outcome distribution. In fact, one interpretation of this $V$ function is as a transformation linking the distribution of $T_{\bx}$ under covariates $\bx$, and the baseline distribution of $T_0$,
\begin{align}
V(T_{\bx}\mid\bx) \sim S_0.
\end{align}
Under this model, the $p$th quantile survival time for subjects under covariate pattern $\bx$ is 
\begin{align}
t^{(p)}_{\bx} & = S^{-1}(p \mid \bx) = V^{-1}(S_0^{-1}(p) \mid \bx).
\end{align}
Now it may no longer be the case that the ratio of $p$th quantile survival times between covariate patterns $\bx$ and $\bx'$ is a constant factor, but rather a quantile-varying acceleration factor
\begin{align}\label{eq:aftqvaf}
\xi(p\mid\bx,\bx', S_0) = t^{(p)}_{\bx} / t^{(p)}_{\bx'} = \frac{S^{-1}(p \mid \bx)}{S^{-1}(p \mid \bx')} = \frac{ V^{-1}(S_0^{-1}(p) \mid \bx) 	}{ V^{-1}(S_0^{-1}(p) \mid \bx') },
\end{align}
with notation explicitly capturing the additional potential for dependence on $p$ and $S_0$.

\subsubsection{Examples and Interpretation}

To emphasize both the flexibility and interpretability of this new quantity, Figure~\ref{fig:SAF1} shows sample survivor curves and corresponding acceleration factors under simple forms of quantile-varying effect for a single contrast between exposure levels $X=1$ and $X=0$, with baseline $S_0(t) = \exp(-0.3t)$. For simplicity we will interpret the effects at $p=0.75$ and $p=0.25$, i.e., the time by which 25\% and 75\% of people experience the event, respectively.

For reference, the black curve in each figure shows a constant acceleration factor of $\exp(0.5) \approx 1.65$ across quantiles. The red curve shows a protective effect that is increasingly pronounced among later-onset cases, with $\xi(0.75\mid 1, 0)=1.25$ and $\xi(0.25\mid 1, 0)=2$. In words, the estimated time by which 25\% of the exposed die is 1.25 times as great as that among the unexposed, but the estimated time by which 75\% of the exposed die is 2 times greater than unexposed. Conceptually, this form of protective effect corresponds with delayed onset of all cases among the exposed, but specifically a much longer tail of late-onset cases compared to a standard AFT protective effect.

\begin{figure}
\begin{center}
	\minipage{0.5\textwidth}%
	\includegraphics[width=\linewidth]{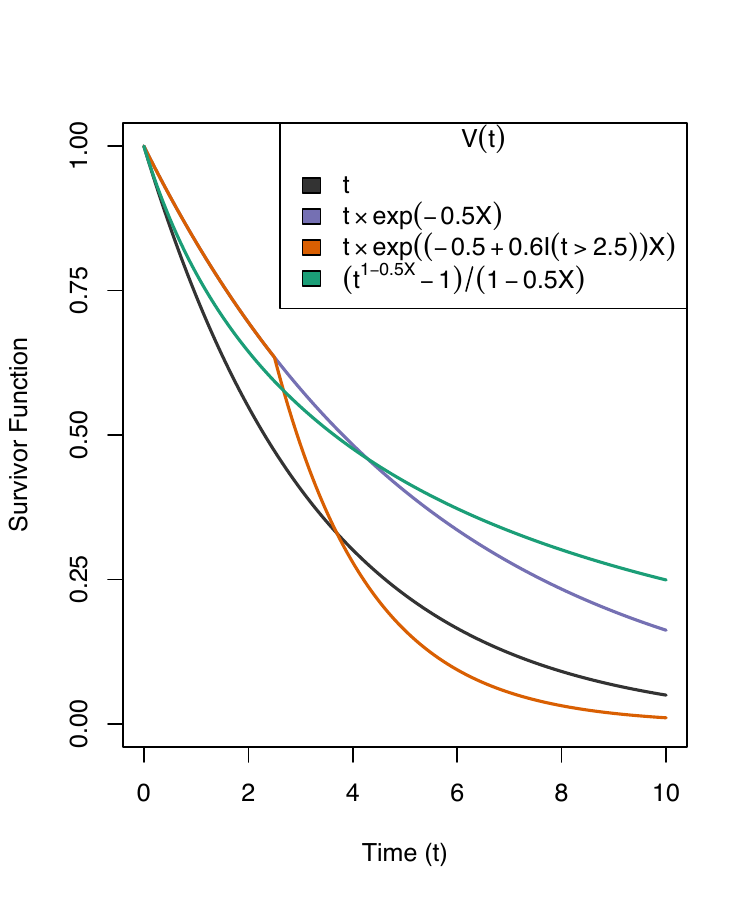}
	\endminipage\hfill
	\minipage{0.5\textwidth}
	\includegraphics[width=\linewidth]{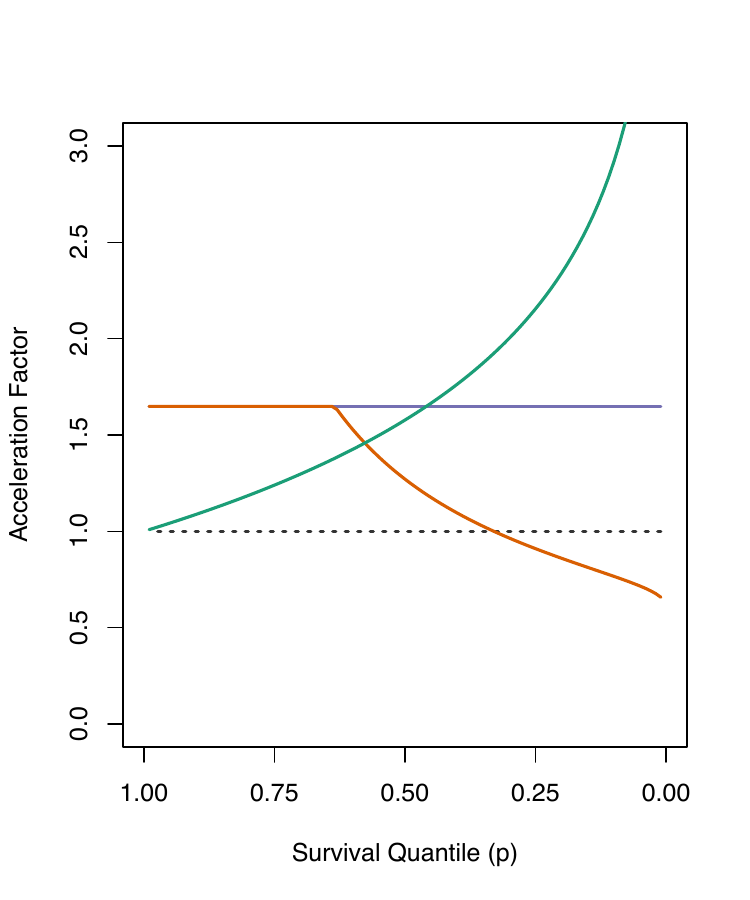}
	\endminipage
	\caption{\textit{Sample survivor curves (left panel) and corresponding, possibly quantile-varying, acceleration factors (right panel). Baseline survivor function shown is} $S_0(t) = \exp(-0.3t)$.\label{fig:SAF1}}
\end{center}
\end{figure}

The blue curve shows a more nuanced effect that delays the earliest cases, while also accelerating later onset cases. Numerically, $\xi(0.75\mid 1, 0)=1.65$ and $\xi(0.25\mid 1, 0)=0.9$, meaning the estimated time by which 25\% of the exposed die is 1.65 times as great as that among the unexposed, but the estimated time by which 75\% of the exposed die is only 0.9 times as great as among the unexposed. Conceptually, this form of effect is a `compressing' of the outcome distribution, with earlier events being delayed and later events being accelerated. This is visible in the relative steepness of blue survivor curve, with more than 50\% of all events occurring between times 2 and 4. Furthermore, this represents an effect with `crossing survivor curves', which despite being common in certain health research domains cannot be modeled by standard proportional hazards or AFT models. In summary, we see that this approach to conceptualizing covariate effects for time-to-event outcomes yields nuanced and interpretable insights beyond what is available from standard proportional hazards or AFT models.

\section{Model Definition}

The proposed quantile-varying AFT model is purposefully general with respect to the baseline survivor distribution $S_0$ and the time-varying covariate process $V$. In this section we outline several choices for specifying these model components, weighing tradeoffs between flexibility, stability, and computation. While this modeling framework in principle admits estimation under both frequentist and Bayesian paradigms, we focus on the latter approach and employ a Markov Chain Monte Carlo (MCMC) estimation routine via the No-U-Turn sampler implemented in the Stan language \citep{carpenter2017stan}.

\subsection{Specification of the covariate process $V$}

Consider a $d$ length vector of baseline covariates $\bX$. For ease of exposition we focus on an exposure of interest $X_1$ specified with a flexible regression effect. However, this can readily extend to allow multiple such exposures of interest. 

The form of the covariate process $V$ dictates the potential shapes the quantile-varying acceleration factor for $X_1$ can take, and requires a balance of flexibility and stability. We focus on spline-based methods, which require a vector of knots $\btau$ characterizing a set of $J$ basis functions $B_1,\dots,B_J$, and corresponding coefficients $\balpha = (\alpha_1,\dots,\alpha_J)\trans$. This results in the specification
\begin{align}\label{eq:Vproc}
V(t\mid\bX) = t\times \exp\left(-\bX\trans\bbeta - X_1\sum_{j=1}^J\alpha_jB_j(t\mid \btau)\right),
\end{align}
Note that when $\balpha = \bzero$, then this reduces to the standard AFT model, allowing straightforward model comparison to assess the flexible effect specification. Then letting $B'_j(t\mid \btau) = d B_j(t\mid \btau)/dt$, the derivative of the covariate process used in likelihood computation has the form
\begin{align}
v(t\mid\bX) = \frac{d}{dt}V(t\mid\btau) = V(t\mid\btau)\left[\frac{1}{t} - X_1\sum_{j=1}^J\alpha_jB'_j(t\mid \btau)\right].
\end{align} 

One specification inspired by the parametric proportional hazards spline model of \citet{royston2002flexible} and discussed by \citet{crowther2022flexible} is the natural cubic spline basis, which combines cubic polynomial basis functions with a restriction that the ends beyond the lower and upper boundary knots be linear. 
Natural cubic spline basis functions $B_k$ and $B'_k$ are available in statistical software, and our implementation uses a set of numerically stable orthogonalized spline bases, from which the corresponding $\beta$ and $\balpha$ are then backtransformed. The resulting $V$ combines flexibility and stability, with the added advantage of being a smooth function of time. 

However, the inverse $V^{-1}(t\mid\bX)$ used in the quantile-varying acceleration factor \eqref{eq:aftqvaf} does not have a closed form, and must be computed numerically. Moreover, this specification does not explicitly enforce the constraint that $V$ be a non-decreasing function, as has been discussed in analogous flexible parametric survival functions using restricted cubic splines \citep{royston2002flexible,crowther2022flexible}. However, as has been repeatedly shown with other such methods in the literature, in our application this specification nevertheless performs well.

An alternative that is both computationally simpler and also enforces the proper constraints is specifying $V$ as a piecewise linear function, yielding a simplified analytical form and closed form inverse. Define $J + 2$ knots $0=\tau_{0} < \tau_{1}<\dots<\tau_{J}<\tau_{J+1}=\infty$, with piecewise linear basis functions
$B_j(t\mid\btau) =t^{-1} (\min\{t,\tau_{j+1}\} - \tau_{j})_{+}$
where $(z)_{+} = \min\{0,z\}$. Then $V$ simplifies to
\begin{align}
V(t\mid\bX) = t\times \exp\left(-\bX\trans\bbeta\right)\left[\sum_{j=1}^{J}\exp\left(-X_1\alpha_j\right) B_{j}(t\mid\btau) \right],
\end{align}
with the straightforward derivative $v(t\mid\bX) = \exp\left\{ -\bX\trans\bbeta - \sum_{j=1}^{J} X_1\alpha_j \bbI(\tau_{j} \leq t < \tau_{j+1}) \right\}$. Computation of the inverse is also straightward, and left to Appendix~\ref{sec:vinvderiv} of the Supplementary Materials. As above, this reduces to the standard AFT model when $\balpha = \bzero$.

\subsection{Specification of the baseline distribution $S_0$}
As with the specification of $V$, there are numerous possible choices of baseline distribution characterizing $S_0$, both fully parametric and semi-parametric. Parametric specifications have several advantages in this setting: they are computationally efficient, well-defined across all quantiles, have tractible inverse survivor functions, and can lead to improved efficiency in smaller samples. Two such parametric specifications are the log-Normal baseline distribution with survivor function defined by $S_0(t\mid \mu, \sigma) =1 - \Phi(\log t - \mu)/\sigma^2$ where $\Phi(\cdot)$ is the standard normal distribution function, and the Weibull baseline distribution defined by $S_0(t\mid \mu, \sigma) = \exp\left\{ \left[t\times\exp(-\mu)\right]^{\sigma} \right\}$. Let $\bphi = (\mu,\sigma)\trans$ denote the collection of parameters corresponding to the baseline distribution.

Nevertheless, an important benefit of the Bayesian paradigm is the well-established literature on semi-parametric AFT survival models with flexible baseline distributions, such as Dirichlet process mixture (DPM) models \citep{lee2017accelerated} and Polya tree priors \citep{hanson2009semiparametric}. Here we consider a transformed Bernstein polynomial (TBP) prior for $S_0$ following \citep{zhou2018unified}, which defines a parametric centering distribution having survivor function $S^{*}_{0}(t\mid \bphi)$ (such as the Weibull or log-Normal defined above), then applies a transformation on the quantile scale using Bernstein polynomial functions to capture a wide array of distributions. Formally, define a vector $\bw$ of length $K$ such that $\sum_{k=1}^K w_k = 1$, and the Beta($a,b$) distribution function $G(p \mid a,b) =  \Gamma(a+b)[\Gamma(a)\Gamma(b)]^{-1} p^{a-1}(1-p)^{b-1}$.
Then the baseline survivor function is 
\begin{align}
S_0(t \mid \bphi, \bw) = \sum_{k = 1}^{K} w_k G(S^{*}_0(t \mid \bphi)\mid k, K - k + 1).
\end{align}
Because the domain of $G$ and the range of $S^{*}_0$ are both [0,1], this represents a flexible spline transformation of the centering parametric distribution on the scale of survival quantiles. In particular, if $\bw = (J^{-1},J^{-1},\dots,J^{-1})\trans$, then $S_0 = S^{*}_0$, so the TBP specification contains the centering parametric model, but can characterize a wide array of survival distribution shapes. An illustration is provided in Appendix~\ref{sec:tbp_detail} of the Supplementary Materials. Finally, we place a $\Dirichlet(\theta)$ prior on $\bw$ with $\theta >0$, where larger values of $\theta$ correspond to tighter concentration of the elements of $\bw$ around $J^{-1}$ and therefore tighter concentration of $S_0$ around $S^{*}_0$.

This specification offers several advantages over other flexible baseline specifications mentioned previously. Importantly, each $G$ function can be computed recursively, so overall computation of $S_0$ is efficient. Moreover, the TBP prior can be sampled using the No-U-Turn algorithm implemented in the Stan language, as described below. By contrast, many other Bayesian non-parametric specifications such as Polya trees and DPM models have discrete parameters that are not admitted in Stan, instead requiring specialized computational methods such as custom MCMC samplers and data augmentation \citep{hanson2009semiparametric,lee2017accelerated}. We believe this is the first such implementation of the TBP prior baseline specification in Stan, allowing us to examine the performance of the specification in a new computational environment.

The main tradeoff with any flexible form for $S_0$ compared to a fully parametric specification is the increased computational cost, both for the sampler as well as the numerical computation of the inverse function $S^{-1}_0$ and associated acceleration factors.

\subsection{Likelihood}

Another important benefit of the Bayesian approach is the ability to seamlessly handle arbitrary censoring and left truncation. Let $(Y^{l},Y^{u})$ the left and right observed endpoints of a censoring interval around a true event time $T$, such that $Y^{l}\leq T \leq Y^{u}$. Right-censoring corresponds with $Y^{u} = \infty$. Define the indicator $\Delta = \bbI(Y^{l} = Y^{u})$ to be a subject observed to experience the event exactly at time $Y^{l}$. Finally, let $L$ represent the possible left-truncation time. Along with the baseline covariates $\bX$, denote the observed data for the $i$th subject $\Dsc_i = \{y_i^{l},y_i^{u},\delta_i,l_i,\bx_i\}$.

After specifying $V$ and $S_0$, denote the full set of parameters $\bpsi = (\bbeta\trans, \balpha\trans, \bphi\trans, \bw\trans)\trans$. Then under non-informative censoring, the $i$th subject's likelihood contribution is
\begin{align}
\Lsc_i(\bpsi\mid \Dsc_i) = \frac{[f_0(V(y_i^{l}\mid \bX_{i}))v(y_i^{l}\mid \bX_{i})]^{\delta_i}[S_0(V(y_i^{l}\mid \bX_{i})) - S_0(V(y_i^{u}\mid \bX_{i}))]^{(1-\delta_i)}}{S_0(V(l_i\mid \bX_{i}))}
\end{align}
where $f_0$ is the density function corresponding to the baseline distribution. By convention, $S_0(\infty) = 0$, so under right-censoring this reduces to the standard censored data likelihood.

\subsection{Bayesian Computation and Prior Specification}

As previously mentioned, we propose Bayesian estimation via the No-U-Turn sampler implemented by the Stan language \citep{carpenter2017stan}. In brief, this MCMC algorithm uses gradient information on the log-posterior to generate Markov transitions that efficiently explore the posterior distribution. The implementation can be easily called from \texttt{R} via the \texttt{rstan} package \citep{standevelopmentteam2020rstan}.

To complete the model specification, we consider priors on the parameters $\bbeta$, $\balpha$, and $\bphi$. The No-U-Turn sampler does not require or leverage conjugacy between prior and posterior, so prior distributions can be chosen or adjusted without changing the implementation of the sampler. In the application below, we adopt flat priors for regression parameters $\bbeta$ and $\balpha$. For parametric distributions, we also adopt a flat prior for the log location parameter $\log \mu$, and for the scale parameter a $\sigma \sim \text{Gamma}(a_{\sigma},b_{\sigma})$ prior. Under the TBP prior, we instead follow previous literature by specifying a multivariate normal prior distribution on $\log\mu$ and $\log\sigma$ centered at 0 with covariance $\Sigma_{\bphi}$. We adopt a $\bw \sim \text{Dirichlet}(\theta)$ prior for the weights, and a $\theta \sim \text{Gamma}(a_{\theta},b_{\theta})$ hyperprior on $\theta$, regulating the level of flexibility around the parametric centering distribution.

\subsection{Model Evaluation and Comparison}

A conceptual benefit of our proposed modeling framework is that the flexible structures naturally encompass simpler models: the standard AFT model is nested within the flexible effect specification of covariate process $V$, and a fully parametric baseline is nested within the TBP prior for $S_0$. In this section, we propose a model evaluation metric to inform decisions regarding the necessary level of model complexity via the \texttt{loo} package in \texttt{R} \citep{vehtari2017practical}.

The expected log pointwise predictive density (ELPD) is a metric for how well a fitted model can predict future out-of-sample data, with larger values meaning better predictive ability. For $n$ future observations $\ytilde_1,\dots,\ytilde_{n}$, it is defined via the posterior predictive density $p(\ytilde\mid\Dsc)$ as
\begin{align}
\text{ELPD} = \sum_{i=1}^n \int p(\ytilde_i)\log p(\ytilde_i\mid\Dsc)d\ytilde_i.
\end{align}
While typically future out-of-sample data is not available, the ELPD can be estimated by leave-one-out cross validation by averaging the log posterior predictive distribution for each observed data point of a model fit excluding that data point. This quantity can in turn be estimated efficiently from a single Bayesian model fit via Pareto smoothed importance sampling, which we denote $\widehat{\text{ELPD}}_{\text{psis-loo}}$ \citep{vehtari2017practical}, and has shown improved performance relative to other common Bayesian model criteria, such as Deviance Information Criterion (DIC).

Alternatively, in Appendix~\ref{sec:bayesfactor} of the Supplementary Materials we derive a Bayes factor for evidence of a quantile-varying effect against the constant effect encoded when $\balpha=\bzero$.

\subsection{Computation of Regression Standardized Acceleration Factors}

Importantly, under the covariate process $V$ defined by \eqref{eq:Vproc}, the quantile-varying acceleration factor \eqref{eq:aftqvaf} depends on the specified values of all covariates $\bx$ and $\bx'$, not just those that differ. This conditionality on the values of all covariates may be insightful if interest is in assessing effect heterogeneity in subpopulations defined by specific covariate patterns. However, practical interest is often in assessing the effect of an exposure in a population standardized with respect to the other covariates. Therefore, in this section we propose a regression standardization approach to estimating the quantile-varying acceleration factor for a particular covariate of interest, averaged over the distribution of other covariates. Conceptually, the goal is to first estimate the survivor curves we would observe in the population if everyone was alternately exposed or unexposed, and then back out the quantile-varying acceleration factor that relates the two curves. 

For clarity, consider a single binary exposure of interest $X$, and vector of additional covariates $\bZ$. Then the marginal ratio of interest is

\begin{align}
\xi(p\mid X=1,X'=0) = \frac{ S^{-1}(p \mid X=1) }{ S^{-1}(p \mid X=0)}.
\end{align}

Following \citet{rothman2021modern} and \citet{sjolander2016regression}, define the survivor function for $X=x$, standardized to the distribution of $\bZ$, as 
\begin{align}
S_{\bZ}(t \mid x) = E_{\bZ}[P( T > t \mid X=x, \bZ)].
\end{align}
We then define the corresponding standardized quantile-varying acceleration factor as
\begin{align}
\xi_{\bZ}(p\mid X=1,X'=0) = \frac{ [S_{\bZ}]^{-1}(p \mid X=1) }{ [S_{\bZ}]^{-1}(p \mid X=0) }.
\end{align}
where $[S_{\bZ}]^{-1}(p\mid X=x)$ is the function solving $S_{\bZ}(t \mid X=x) - p = 0$ for $t$. This contrast represents the magnitude of the horizontal shift in the standardized survivor curve $S_{\bZ}$ between $X=1$ and $X=0$, at each quantile $p$.

To estimate and quantify uncertainty for these contrasts, we develop a novel approach based on the Bayesian g-formula \citep{keil2018bayesian}. In brief, for each MCMC draw $m=1,\dots,M$, for each $X=x$ we compute the standardized survivor function 
\begin{align}
S^{(m)}_{\bZ}(t\mid X = x) = n^{-1}\sum_{i=1}^n S_{\bZ}(t\mid X = x,\bZ=\bz_i;\bpsi^{(m)}),
\end{align}
and then form contrast of interest
\begin{align}
\xi^{(m)}_{\bZ}(p\mid X=1,X'=0) = \frac{[S^{(m)}_{\bZ}]^{-1}(p\mid X = 1)}{[S^{(m)}_{\bZ}]^{-1}(p\mid X = 0)}.
\end{align}
This may require numerical evaluation of the inverse standardized survivor functions. Estimating posterior mean and credible intervals of $\xi_{\bZ}$ proceeds using the mean and quantiles of $\xi^{(1)}_{\bZ},\dots,\xi^{(M)}_{\bZ}$.

\subsection{Simulation Study}\label{sec:sims}

To examine the operating characteristics of the proposed AFT framework, we performed a simulation study detailed in Appendix~\ref{sec:appsims} of the Supplementary Materials that illustrates the performance of quantile-varying acceleration factor estimates across several settings and model specifications. As data generating mechanisms, we considered true log-Normal baseline hazards having a binary covariate with either constant or quantile-varying effects, and two nuisance covariates.

We fit log-Normal and Weibull parametric models as well as a Weibull-centered TBP prior model, each with either constant, piecewise, or spline effect specifications. This allowed the assessment of performance when the baseline and/or effect were either correctly specified, flexibly approximated, or incorrectly specified. We compared the estimated ELPD metric, the integrated squared error of the baseline hazard $\int (\widehat{h}_0(t) - h^*_0(t))^2dt$, and the bias, standard error, and coverage probability of the acceleration factors both conditionally and after regression standardization.

With details in the Supplementary Materials, models with correctly specified effects as well as correctly specified baselines had the best ELPD and least biased, most precise estimates. The next-best performing models were those with TBP prior baselines, which nevertheless exhibited slightly elevated bias and mild undercoverage of the credible intervals especially in the lower quantiles where information loss due to censoring was very high. Models with Weibull parametric baselines performed poorest across all metrics. We also observed that the TBP prior model implemented in Stan could have trouble initializing in a minority of simulation iterations, which may be due to a combination of finite-difference gradient computations used in the No-U-Turn sampling algorithm with numerical instability that may arise in computation of the TBP survivor function. This is discussed further in the Supplementary Materials.

Practically, these results support the use of ELPD model diagnostic criteria to select the final model specification in the data application below, as it was generally concordant with estimation performance of both the baseline distribution and acceleration factors.

\section{Application: Cohort Study of Incident AD and Dementia}
Motivating the proposed AFT model is the study of adverse cognitive outcomes among older adults, as long timescales and complex disease etiology lend themselves to considering flexible covariate effects on the quantile scale. In this section we investigate risk factors for AD and dementia in older adults using data collected by the Religious Orders Study and Memory and Aging Project (ROSMAP) cohort studies ongoing since 1994 and 1997 respectively \citep{bennett2018religious}. Our analysis focuses on flexible estimation of the association of the genetic marker APOE-$\epsilon4$ with the timing of AD or dementia onset. Previous analyses of similar cohorts have simply compared incidence rates within age categories to examine whether this marker had differential effects through time \citep{kukull2002dementia}. So, estimating a quantile-varying acceleration factor for APOE-$\epsilon4$ is of clinical relevance, while also accounting for other risk factors.

2694 subjects were enrolled without AD or dementia between ages 65 and 86, and followed until withdrawal or death. Subjects underwent cognitive screening annually to diagnose onset of AD or dementia, and death status was monitored continuously. Table~\ref{tab:baseline} summarizes a set of baseline binary risk factors collected on the subjects: marital status at baseline, sex, education level, race/ethnicity, and presence of the APOE-$\epsilon4$ genetic variant. The final analysis dataset includes 2335 subjects with complete baseline information. The outcome is defined by the time of diagnosis of AD or dementia, with death treated as a censoring mechanism, yielding a cause-specific analysis. Because we only include subjects with age at least 65, the time scale of analysis is ``years since age 65.'' Importantly, our analysis accounts for the presence of left truncation (or ``delayed entry'') by subjects who enroll after age 65. Though this framework extends to interval censoring, given the short visit intervals relative to the timescale, for this analysis we defined right censoring times for AD/dementia onset at the midpoint of the corresponding visit interval.

\begin{table}
\centering
\caption{\textit{Baseline covariates by observed AD/dementia and death outcome status.}\label{tab:baseline}}{\resizebox{1.0\textwidth}{!}{\begin{tabular}{rccccc}
  \hline
 & 	     & Censored prior & AD/dementia & Death & AD/dementia \\ 
 & 	     & to AD/dementia & and censored & without & diagnosis \\ 
 & Total (\%)  & or death (\%)  & prior to death (\%)  & AD/dementia (\%)  & and death (\%)  \\ 
\hline
Total & 2335 (100\%) & 750 (100\%) & 123 (100\%) & 891 (100\%) & 571 (100\%) \\  
  \hline
  White Race/Ethnicity & 2178 (93.3\%) & 687 (91.6\%) & 100 (81.3\%) & 849 (95.3\%) & 542 (94.9\%) \\ 
  Male Sex & 648 (27.8\%) & 147 (19.6\%) & 24 (19.5\%) & 316 (35.5\%) & 161 (28.2\%) \\ 
  Married at Study Entry & 462 (19.8\%) & 216 (28.8\%) & 29 (23.6\%) & 145 (16.3\%) & 72 (12.6\%) \\ 
  $\geq 15$ Years of Education & 1621 (69.4\%) & 532 (70.9\%) & 80 (65\%) & 609 (68.4\%) & 400 (70.1\%) \\ 
  APOE-$\epsilon4$ Genetic Variant & 575 (24.6\%) & 156 (20.8\%) & 53 (43.1\%) & 173 (19.4\%) & 193 (33.8\%) \\ 
   \hline
\end{tabular}}}
\end{table}

We compare the fits of standard AFT models with those having piecewise and spline forms for $V$, under Weibull and log-Normal baseline specifications as well as a TBP prior baseline with $K=5$, centering around the Weibull distribution. We set 4 break points for the piecewise linear effect at 7.5 year intervals across the follow up period, and for the spline effect we set 2 internal knots at quantiles on the log scale. The difference between these specifications is due to the spline being naturally more flexible, allowing it to smooth across knots with irregular spacing, while the piecewise linear model requires break points that span the follow-up period to achieve flexibility. For the parametric scale parameters we set $\sigma \sim \text{Gamma}(0.3,0.05)$, having prior median 1.46 and 95\% central mass between 6e-5 and 38. For the TBP baseline specification, we set a hyperprior of $\theta \sim \text{Gamma}(1,1)$, and consistent with the literature on this model to minimize confounding between the flexibility of $\bphi$ and $\bw$, adopted a prior covariance $\Sigma_{\bphi}$ for the Weibull centering parameters equal to the maximum likelihood estimate from a standard Weibull AFT scaled by 10. For each model we ran three chains each for 2000 adaptation iterations and 10000 samples, totalling 30000 samples. After sampling, all potential scale reduction factors were below $1.01$ and trace plots indicated good mixing. For comparison, we include results from a standard Frequentist Cox proportional hazards model. Additionally, we present Bayesian proportional hazard and proportional odds comparators fit with TBP prior baseline hazards in Appendix~\ref{sec:addldata} of the Supplementary Materials.

Table~\ref{tab:ad_results} reports the estimates of regression parameters across all AFT specifications, as well as frequentist results from a Cox proportional hazards model. For the AFT models, positive estimates of $\bbeta$ correspond with delayed onset of AD or dementia, as do negative estimates for the Cox model. The coefficients estimated for white race/ethnicity, marital status, female sex, and education are stable across all model specifications. Interpreting the Weibull AFT with constant effect of APOE-$\epsilon 4$, for example, indicates that being married is associated with a $\exp(0.09)=1.09 $ times greater median time to onset of AD or dementia, with 95\% credible interval of (1.02,1.19). Flexible effect coefficients of APOE-$\epsilon 4$ cannot be directly interpreted on the quantile scale, therefore we present graphical tools below.

\begin{table}[ht] 
\centering
\caption{\textit{Regression estimates for time to onset of AD or dementia in the absence of death. AFT results are posterior medians and 95\% credible intervals for regression parameters. Cox model results are log-hazard ratio estimates and 95\% confidence intervals.} \label{tab:ad_results}}{\resizebox{\textwidth}{!}{\begin{tabular}{lcccc}
  \hline
& & \multicolumn{3}{c}{AFT Model} \\ \cline{3-5}
  & Cox PH & log-Normal & Weibull & TBP (Weibull Centered) \\ 
  \hline
White Race/Ethnicity, $\beta_1$ &  &  &  &  \\ 
~~Constant & -0.28 (-0.57, 0.01) & 0.18 (0.03, 0.31) & 0.08 (-0.03, 0.19) & 0.08 (-0.02, 0.19) \\ 
~~Piecewise Linear &  & 0.18 (0.05, 0.3) & 0.08 (-0.02, 0.17) & 0.07 (-0.03, 0.16) \\ 
~~Restricted Cubic Spline &  & 0.15 (0.03, 0.28) & 0.07 (-0.02, 0.16) & 0.07 (-0.02, 0.16) \\ 
  Male Sex, $\beta_2$ &  &  &  &  \\ 
~~Constant & 0.06 (-0.11, 0.23) & -0.04 (-0.13, 0.04) & -0.02 (-0.08, 0.05) & -0.02 (-0.08, 0.04) \\ 
~~Piecewise Linear &  & -0.05 (-0.12, 0.03) & -0.02 (-0.08, 0.04) & -0.02 (-0.08, 0.03) \\ 
~~Restricted Cubic Spline &  & -0.04 (-0.11, 0.03) & -0.02 (-0.07, 0.03) & -0.02 (-0.07, 0.03) \\ 
  Married at Study Entry, $\beta_3$ &  &  &  &  \\ 
~~Constant & -0.26 (-0.47, -0.04) & 0.13 (0.03, 0.23) & 0.1 (0.02, 0.19) & 0.1 (0.03, 0.18) \\ 
~~Piecewise Linear &  & 0.13 (0.04, 0.22) & 0.09 (0.02, 0.16) & 0.08 (0.02, 0.16) \\ 
~~Restricted Cubic Spline &  & 0.13 (0.04, 0.22) & 0.09 (0.02, 0.16) & 0.08 (0.02, 0.15) \\ 
  $\geq$15 Years of Education, $\beta_4$ &  &  &  &  \\ 
~~Constant & -0.1 (-0.26, 0.07) & 0.07 (-0.01, 0.16) & 0.04 (-0.02, 0.1) & 0.03 (-0.03, 0.1) \\ 
~~Piecewise Linear &  & 0.07 (0, 0.15) & 0.03 (-0.02, 0.09) & 0.03 (-0.02, 0.08) \\ 
~~Restricted Cubic Spline &  & 0.06 (-0.01, 0.14) & 0.03 (-0.02, 0.09) & 0.03 (-0.02, 0.08) \\ 
  APOE-$\epsilon4$ Genetic Variant, $\beta_5$ &  &  &  &  \\ 
 ~~Constant & 0.76 (0.61, 0.92) & -0.42 (-0.51, -0.34) & -0.28 (-0.35, -0.22) & -0.28 (-0.35, -0.21) \\ 
~~Piecewise Linear &  & -0.79 (-0.95, -0.62) & -0.75 (-0.92, -0.55) & -0.76 (-0.94, -0.53) \\ 
~~Restricted Cubic Spline &  & -2.54 (-2.98, -1.95) & -2.38 (-3.08, -1.23) & -2.31 (-3.10, -0.93) \\ 
  APOE-$\epsilon4$ Genetic Variant, $\alpha_1$ &  &  &  &  \\ 
~~Constant &  &  &  &  \\ 
~~Piecewise Linear &  & 0.86 (0.49, 1.23) & 0.78 (0.35,1.20) & 0.78 (0.28, 1.24) \\ 
~~Restricted Cubic Spline &  & 1.51 (1.12, 1.85) & 1.51 (0.77, 2.01) & 1.49 (0.62, 2.05) \\ 
  APOE-$\epsilon4$ Genetic Variant, $\alpha_2$ &  &  &  &  \\ 
~~Constant &  &  &  &  \\ 
~~Piecewise Linear &  & 0.52 (0.23, 0.80) & 0.74 (0.39,1.06) & 0.8 (0.42, 1.15) \\ 
~~Restricted Cubic Spline &  & 3.83 (2.73, 4.47) & 3.63 (1.45, 4.76) & 3.48 (0.87, 4.79) \\ 
  APOE-$\epsilon4$ Genetic Variant, $\alpha_3$ &  &  &  &  \\ 
~~Constant &  &  &  &  \\ 
~~Piecewise Linear &  & 0.46 (0.11, 0.81) & 0.97 (0.59,1.34) & 1.01 (0.59, 1.41) \\ 
~~Restricted Cubic Spline &  & 1.07 (0.71, 1.40) & 1.34 (0.77, 1.77) & 1.33 (0.67, 1.81) \\ 
  APOE-$\epsilon4$ Genetic Variant, $\alpha_4$ &  &  &  &  \\ 
~~Constant &  &  &  &  \\ 
~~Piecewise Linear &  & -0.38 (-1.06, 0.45) & 0.41 (-0.23,1.20) & 0.39 (-0.34, 1.23) \\ 
~~Restricted Cubic Spline &  &  &  &  \\ 
\hline
\end{tabular}}}
\end{table}

The top panel of Table~\ref{tab:diagnostics} compares estimates of ELPD model criterion for each AFT model. In each case, the spline and piecewise-linear effect specifications outperformed the standard AFT specification. The log-Normal models uniformly underperformed, while the Weibull and TBP models performed comparably. To graphically assess the effect of APOE-$\epsilon 4$ we report the TBP model, and present results for other specifications in Appendix~\ref{sec:addldata} of the Supplementary Materials. Results were qualitatively similar for all baseline distributions, with the largest differences in acceleration factor only occurring in the lowest quantiles extrapolated beyond the observed data.

\begin{table}[ht] 
\centering
\caption{Estimated expected log predictive density (ELPD), multiplied by -2 to replicate scale of information criteria. Smaller values indicate better model fit.
\label{tab:diagnostics}}{\begin{tabular}{lccc}
  \hline
& \multicolumn{3}{c}{AFT Model} \\
\cline{2-4}
 & log-Normal & Weibull & TBP (Weibull Centered)  \\ 
  \hline
\multicolumn{4}{l}{\textbf{AD/Dementia Onset (Death as a Censoring Mechanism)}} \\   
~~Constant & 5862.0 & 5806.3 & 5800.4 \\ 
~~Piecewise Linear & 5841.5 & 5788.4 & 5782.2 \\ 
~~Restricted Cubic Spline & 5814.9 & 5780.9 & 5776.8 \\ 
\multicolumn{4}{l}{\textbf{Death (AD/Dementia as a Time-Varying Covariate)}} \\   
~~Constant & 9997.2 & 9666.7 & 9629.8 \\ 
~~Piecewise Linear & 9919.8 & 9636.7 & 9601.7 \\ 
~~Restricted Cubic Spline & 9884.5 & 9600.9 & 9566.7 \\ 
\hline
\end{tabular}}
\end{table}

Figure~\ref{fig:ad_marg_surv_af_tbp} shows the estimated survivor functions and corresponding quantile-varying acceleration factors for the APOE-$\epsilon4$ genetic variant, after regression standardization over the distribution of the other baseline covariates. These figures confirm other findings that APOE-$\epsilon4$ is associated with earlier onset of AD and dementia. However, quantile-varying effects also indicate that the acceleration is strongest among the earliest cases and subsequently diminishes. Both piecewise and spline models estimate that the time by which the first 10\% of those living with APOE-$\epsilon4$ develop AD or dementia is earlier than those without the variant by a factor of about 0.5; the median times by which people develop AD or dementia differ by a factor of about 0.75, and the times by which 75\% develop AD or dementia differ by a factor of about 0.85. Due to censoring of those with advanced age, the acceleration factor at lower quantiles reflects parametric extrapolation beyond the observed distribution, represented in the figure by grey shading. Nevertheless, this finding has clear clinical significance, indicating the particular need to monitor for early onset AD/dementia at younger ages among those with APOE-$\epsilon4$.

\begin{figure}[ht] 
\begin{center}
	\minipage{0.5\textwidth}%
		\includegraphics[width=\linewidth]{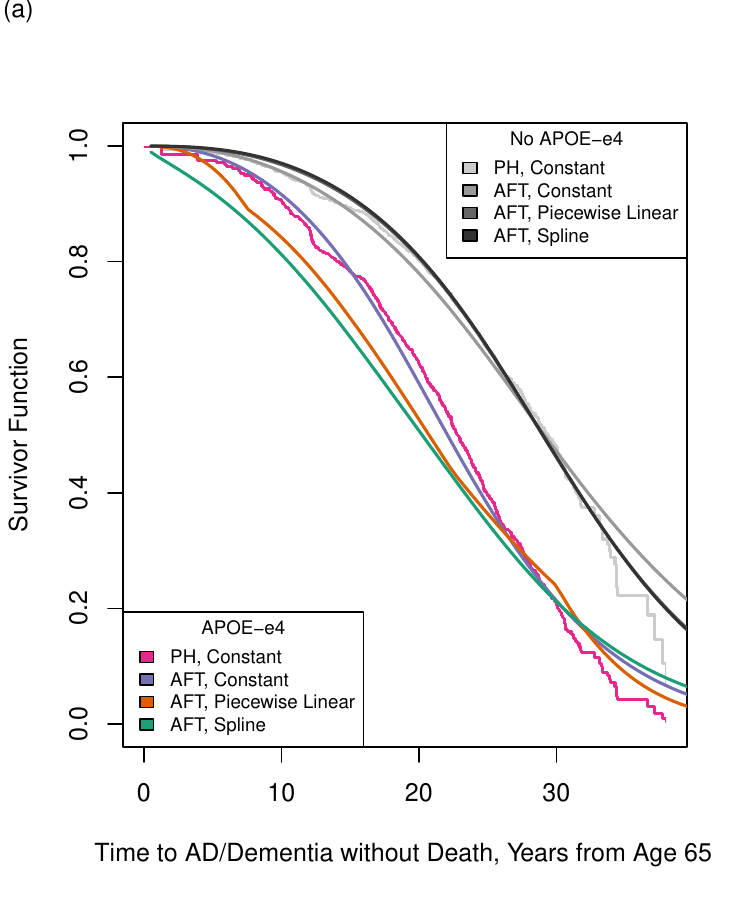}		\endminipage\hfill
	\minipage{0.5\textwidth}
	\includegraphics[width=\linewidth]{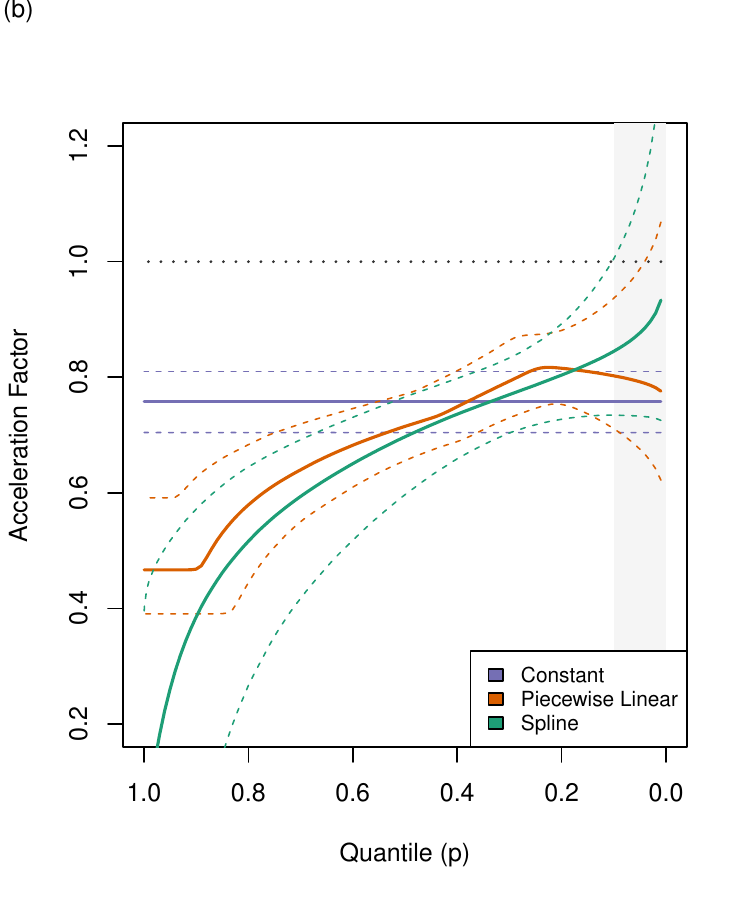}
	\endminipage
\end{center}
\caption{\textit{Under a Weibull-centered TBP prior baseline specification: (a) regression standardized survivor function estimates for onset of AD or dementia without death, averaged over other covariates. Regression standardized estimate from Cox proportional hazards model shown for comparison; (b) regression standardized quantile-varying acceleration factor estimates for onset of AD or dementia without death, averaged over other covariates. 95\% credible intervals represented with dashed lines. Grey shaded region represents area of parametric extrapolation beyond quantiles observed in both groups.} \label{fig:ad_marg_surv_af_tbp}}
\end{figure}

\section{Effects of Time-varying Covariates on the Quantile Scale}

In this section, we extend the proposed AFT model to incorporate binary time-varying covariates, and provide graphical tools for effectively interpreting corresponding effects on the quantile scale. 

To focus on intuition, consider a single time-varying covariate denoted $X_1(t)$ with constant regression effect $\beta_1$. In particular, let $X_1(t)$ be a binary-valued step function, such as an indicator for whether a non-terminal event has occurred by time $t$. Formally, define $X_1(t) = \bbI(t > t_{X})$, where $t_{X}$ is the time at which $X_1$ changes. To simplify notation, consider a single additional covariate time-invariant covariate $X_2$, though inclusion of multiple additional covariates is straightforward. Embedding these covariates directly in the structure for $V$ given by \eqref{eq:Vproc} and setting $\balpha = \bzero$ to denote a constant effect yields 
\begin{equation} \label{eq:tvVproc}
\begin{aligned}
V(t\mid\bX(t)) & = t\times \exp\left( - X_1(t)\beta_1 - X_2\beta_2\right) 
\\ & =  \exp\left(-X_2\beta_2\right) \left[ \min(t,t_{X}) + (t - t_{X})_{+} \exp\left(-\beta_1\right) \right].
\end{aligned}
\end{equation}
With complete derivation given in Appendix~\ref{sec:tvcovderiv} of the Supplementary Materials,  the acceleration factor at quantile $p$ between two subjects depends on each person's value of $X_2$, the change time $t_{X}$ for $X_1$, and the baseline distribution $S_0$. In particular, for those with $X_2=x_2$, the acceleration factor at quantile $p$ for experiencing $X_1$ at $t_X$ versus not experiencing $X_1$ is
\begin{align}\label{eq:afttvaf}
\frac{t_{X}}{S_{0}^{-1}(p)\exp(x_2\beta_2)} + \exp(\beta_1)\left[ 1 - \frac{t_{X}}{S_{0}^{-1}(p)\exp(x_2\beta_2)} \right].
\end{align}
This is a weighted average between 1 and $\exp(\beta_1)$, with weight inversely proportional to the duration from $t_{X}$ to the $p$th quantile survival time in the comparison group, $S_{0}^{-1}(p)\exp(x_2\beta_2)$. Intuitively, before $t_{X}$ there is no difference between the individuals, so the acceleration factor is 1, and then after $t_{X}$ the effect of $X_1$ starts accumulating, and the acceleration factor increasingly shifts towards $\exp(\beta_1)$ as $p$ extends towards 0. Figure~\ref{fig:deathtv_marg_surv_af_tbp} illustrates this dynamic.

Finally, a flexible effect for $X_1(t)$ can also be specified by adapting the form of \eqref{eq:tvVproc}, yielding
\begin{align}\label{eq:tvVprocflex}
V(t\mid\bX(t)) = e^{-X_2\beta_2} \left[ \min\{t,t_{X}\} + (t - t_{X})_{+} \exp\left(-\beta_1 - \sum_{k=1}^K\alpha_kB_k(t - t_{X} \mid \btau)\right) \right].
\end{align}
Following \citet{haneuse2008separation}, this specification characterizes flexibility in the effect of $X_1$ over the time scale $t-t_{X}$ denoting time since the non-terminal event, rather than on the overall time scale of $t$, enabling evaluation of the temporal effect of $X_1$ on its own timescale. Practically, this means that basis functions and knots $\btau$ must be specified on the corresponding time scale.

\subsection{Effect of Incident AD and Dementia on Mortality}
To illustrate the AFT framework with a time-varying binary covariate, we perform a secondary analysis of the cohort study to evaluate the association between onset of AD/dementia and subsequent time to death. We fit models specifying onset of AD/dementia as a binary time-varying covariate, adjusting for the same time-invariant baseline covariates as in the above analysis (including a constant effect for APOE-$\epsilon$4).

For the piecewise linear effect, we set break points at 1, 2, 3, 5, and 10 years after time of AD onset, and for the spline effect we set 2 internal knots at observed quantiles of time from AD onset to death on the log scale. Other settings were as above, though for computation of the acceleration surface described below, we thinned the samples by a factor of 10 to facilitate computation. Table~\ref{tab:deathtv_results} in Appendix~\ref{sec:addldata} of the Supplementary Materials reports estimated model parameters varying baseline survival distribution and effect specification. As before, the estimated baseline covariate coefficients are stable across time-varying effect specifications.

Panels (a) and (c) of Figure~\ref{fig:deathtv_marg_surv_af_tbp} show estimated regression standardized survivor curves under a TBP prior baseline comparing those without AD/dementia onset to those with onset at age 70 and 85, respectively. The curves are identical up until the time of onset, and then once AD/dementia onset occurs mortality increases substantially. The plots indicate similarity between models fit with piecewise and spline effects of AD/dementia onset relative to a constant effect, though the flexible models indicate a small delay in the mortality increase from the time of AD/dementia onset. Corresponding acceleration factors are given in panels (b) and (d), illustrating the trajectory derived in \eqref{eq:afttvaf}, where no association exists before the quantile of AD/dementia onset, followed by an increasingly pronounced association after AD/dementia onset.

\begin{figure}
\begin{center}
	\minipage{0.49\textwidth}%
		\includegraphics[width=\linewidth]{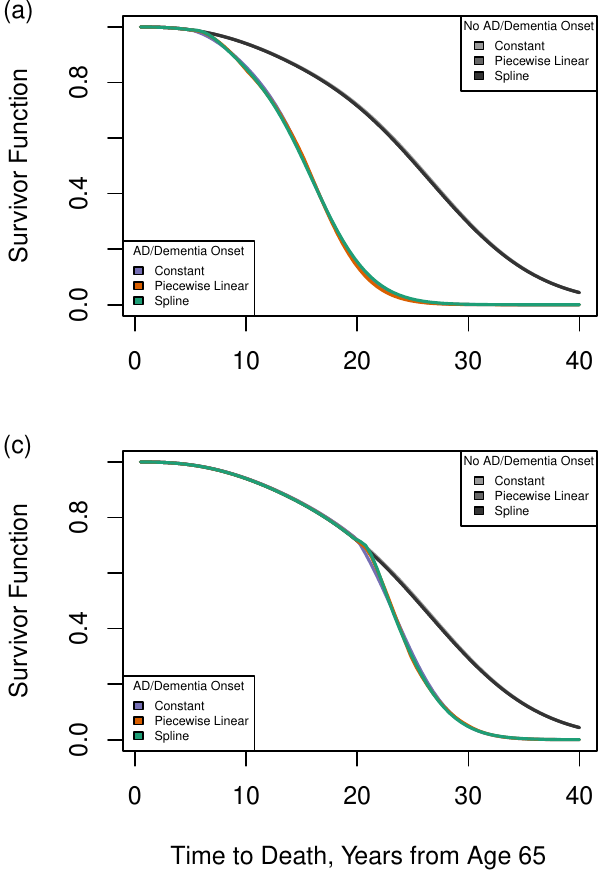}		\endminipage\hfill
	\minipage{0.49\textwidth}
	\includegraphics[width=\linewidth]{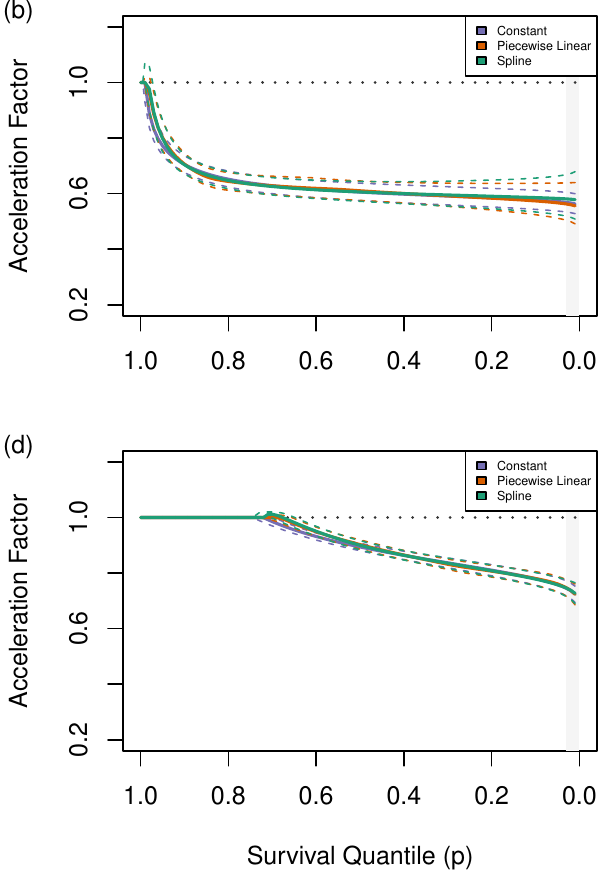}
	\endminipage
\end{center}
\caption{\textit{Under a Weibull-centered TBP prior baseline specification: regression standardized survivor function estimates for mortality following onset of AD or dementia at (a) age 70 and (c) age 85, averaged over other covariates; regression standardized survivor function estimates for mortality following onset of AD or dementia at (b) age 70 and (d) age 85, averaged over other covariates. 95\% credible intervals represented with dashed lines. Grey shaded region represents area of parametric extrapolation beyond quantiles observed in both groups.} \label{fig:deathtv_marg_surv_af_tbp}}
\end{figure}

Plotting acceleration factors for a few AD/dementia onset times of interest may be sufficient in some settings, but does not communicate the quantile-varying effect across the entire range of the time-varying covariate. Instead, Figure~\ref{fig:deathtv_marg_afsurface_tbp_invariant} reports the acceleration factor surface as a contour plot, with the time of AD/dementia onset on the y-axis, the survival quantile on the x-axis, and the color representing the magnitude of the acceleration factor. The two acceleration factor plots in Figure~\ref{fig:deathtv_marg_surv_af_tbp} correspond with cross-sections of this surface, by drawing horizontal lines at times 5 and 20 on the y-axis. More generally, looking horizontally across this plot shows the quantile varying acceleration factor corresponding with different times of AD/dementia onset. However, this plot can also be read vertically, to show how the acceleration factor for a particular quantile changes depending on the timing of the time-varying covariate. For example, drawing a vertical line from 0.5 on the x-axis shows the acceleration factor for median survival, varying across times of AD/dementia onset. Therefore, this single plot conveys complex regression effects both as a function of the survival quantile, as well as of the timing of the time-varying covariate.

\begin{figure}[ht]
\begin{center}
	\minipage{0.33\textwidth}%
		\includegraphics[width=\linewidth]{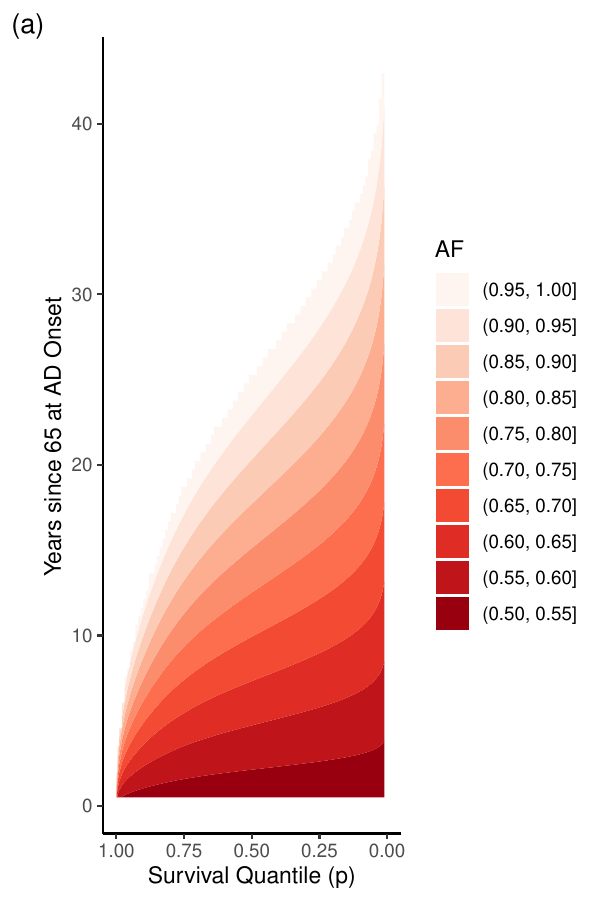}		\endminipage\hfill
	\minipage{0.33\textwidth}
	\includegraphics[width=\linewidth]{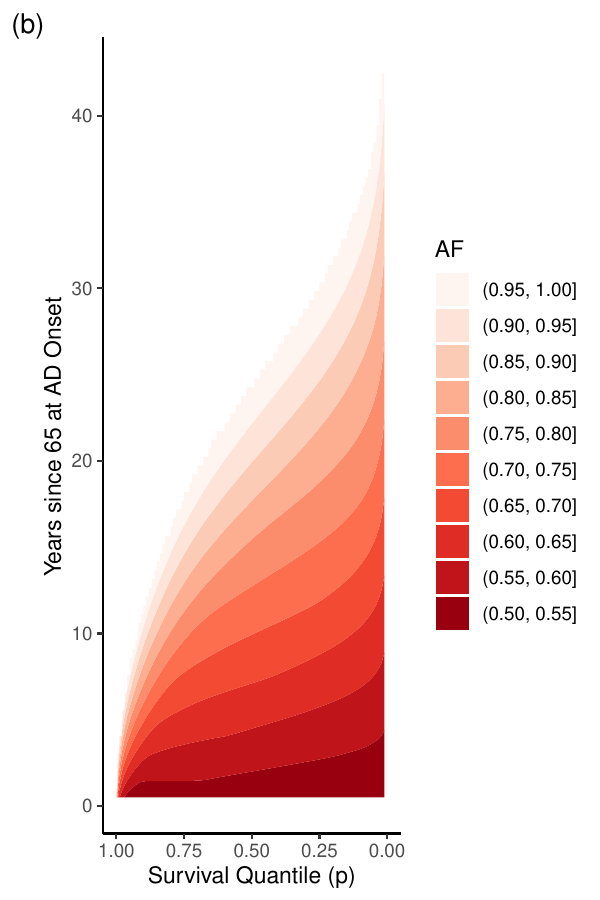}		\endminipage\hfill
	\minipage{0.33\textwidth}
	\includegraphics[width=\linewidth]{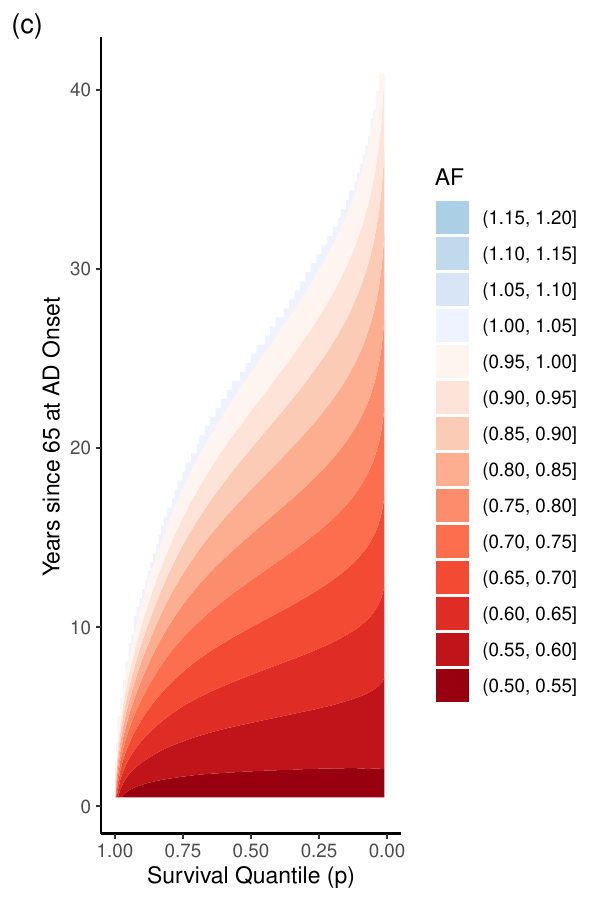}	
	\endminipage
\end{center}
\caption{\textit{Under a Weibull-centered TBP prior baseline specification, contour plots of regression standardized acceleration factor surface estimates for death following onset of AD/dementia, averaged over other covariates. Time of AD/dementia onset is shown on y-axis, and subsequent survival quantile is shown on x-axis. Color indicates acceleration factor at the given survival quantile, comparing those with AD/dementia onset at the specified time and those without AD/dementia. Horizontal cross-sections illustrate quantile-varying acceleration factor for AD/dementia onset at a particular time, while vertical cross-sections illustrate acceleration factor at a particular quantile across times of AD/dementia onset. (a) constant effect specification; (b) piecewise linear effect specification; and (c) spline effect specification.}\label{fig:deathtv_marg_afsurface_tbp_invariant}}
\end{figure}

\section{Discussion}

The AFT model's specification of multiplicative covariate effects on the quantile scale provides an interpretable and attractive alternative to the standard proportional hazards model. Our proposed extensions to the AFT model enabling quantile-varying acceleration factors, and admitting binary time-varying covariates represent important additions to the standard toolbox for survival analysis. Just as the Cox proportional hazards model benefits from straightforward incorporation of time-varying hazard ratios, the ability to add flexibility to the AFT model regression effects expands the scope of scientific inquiry. Motivated by the study of AD in older adults, we found that the association of the APOE-$\epsilon4$ gene with AD onset varied substantially across quantiles, with earlier-onset cases accelerated the most and later-onset cases the least. 

Moreover, the ability to model, summarize, and communicate the effects of binary time-varying covariates creates new opportunities to examine longitudinal health trajectories. The proposed visualization of these effects as a surface across the covariate timescale and the survival quantiles is particularly valuable, as previous work to incorporate time-varying covariates into AFT models has not focused on communication of effects of time-varying components on the quantile scale \citep{hanson2009semiparametric,zhou2018unified}. In our application, this approach illustrated that the association between AD/dementia onset and subsequent mortality varies substantially both across survival quantiles, and depending on the time of AD/dementia onset.

While in principle any increasing choice of $V$ could be used, our proposed form in \eqref{eq:Vproc} using an exponential link has the benefit of representing multiplicative effects of the covariates in a direct extension of the AFT model. In theory, one could use other forms of $V$, with potentially differing interpretation. For example, one might consider additive forms for $V$ using monotonically-increasing I-splines for the time-varying effect of each covariate, akin to a Generalized Additive Model \citep{hastie1986generalized}. Also, while our focus for clear exposition is on a single quantile-varying effect, in principle additional effects could be specified on other parameters. To address the multiplicity of terms, one might incorporate a prior over the $\balpha$ parameters that induces shrinkage towards $\bzero$, i.e., standard `quantile-invariant' effects.

Finally, our work complements the related literature on censored quantile regression \citep{portnoy2003censored,reich2013bayesian}, which specifies additive effects of covariates on the quantile scale, while our model specifies multiplicative effects on the quantile scale. The biological plausibility or clinical relevance of additive versus multiplicative changes to the survival quantiles depends on the application, so the proposed methodology is a valuable alternative to such methods.

\section*{Funding}

This project was supported by the Eunice Kennedy Shriver National Institute of Child Health and Human Development [grant number F31HD102159 to HTR]. The National Institutes on Aging supported the Religious Orders Study [grant numbers P30AG010161 and R01AG015819] and the Rush Memory and Aging Project [grant number R01AG017917].

\section*{Acknowledgments}

We thank the study participants and staff of the Rush Alzheimer’s Disease Center. ROSMAP resources can be requested at \url{https://www.radc.rush.edu}.

\section*{Software}
\label{sec5}

Software in the form of R and Stan code is available at \url{https://github.com/harrisonreeder/aftquantile}.

\bibliographystyle{apalike}
\bibliography{zoterolibrary2022-09-22}

\begin{thebibliography}{}

\bibitem[Bennett et~al., 2018]{bennett2018religious}
Bennett, D.~A., Buchman, A.~S., Boyle, P.~A., Barnes, L.~L., Wilson, R.~S., and
  Schneider, J.~A. (2018).
\newblock Religious {{Orders Study}} and {{Rush Memory}} and {{Aging Project}}.
\newblock {\em Journal of Alzheimer's Disease}, 64(s1):S161--S189.

\bibitem[Carpenter et~al., 2017]{carpenter2017stan}
Carpenter, B., Gelman, A., Hoffman, M.~D., Lee, D., Goodrich, B., Betancourt,
  M., Brubaker, M., Guo, J., Li, P., and Riddell, A. (2017).
\newblock Stan: {{A Probabilistic Programming Language}}.
\newblock {\em Journal of Statistical Software}, 76(1).

\bibitem[Cox and Oakes, 1984]{cox1984analysis}
Cox, D.~R. and Oakes, D. (1984).
\newblock {\em Analysis of Survival Data}.
\newblock Monographs on Statistics and Applied Probability. {Chapman and Hall},
  {London ; New York}.

\bibitem[Crowther et~al., 2022]{crowther2022flexible}
Crowther, M.~J., Royston, P., and Clements, M. (2022).
\newblock A flexible parametric accelerated failure time model and the
  extension to time-dependent acceleration factors.
\newblock {\em Biostatistics}, page kxac009.

\bibitem[Haneuse et~al., 2008]{haneuse2008separation}
Haneuse, S. J.-P.~A., Rudser, K.~D., and Gillen, D.~L. (2008).
\newblock The separation of timescales in {{Bayesian}} survival modeling of the
  time-varying effect of a time-dependent exposure.
\newblock {\em Biostatistics}, 9(3):400--410.

\bibitem[Hanson et~al., 2009]{hanson2009semiparametric}
Hanson, T., Johnson, W., and Laud, P. (2009).
\newblock Semiparametric inference for survival models with step process
  covariates.
\newblock {\em Canadian Journal of Statistics}, 37(1):60--79.

\bibitem[Hastie and Tibshirani, 1986]{hastie1986generalized}
Hastie, T. and Tibshirani, R. (1986).
\newblock Generalized {{Additive Models}}.
\newblock {\em Statistical Science}, 1(3):297--310.

\bibitem[Hern{\'a}n, 2010]{hernan2010hazards}
Hern{\'a}n, M.~A. (2010).
\newblock The hazards of hazard ratios.
\newblock {\em Epidemiology}, 21(1):13--15.

\bibitem[Hsieh, 2001]{hsieh2001heteroscedastic}
Hsieh, F. (2001).
\newblock On heteroscedastic hazards regression models: Theory and application.
\newblock {\em Journal of the Royal Statistical Society: Series B (Statistical
  Methodology)}, 63(1):63--79.

\bibitem[Keil et~al., 2018]{keil2018bayesian}
Keil, A.~P., Daza, E.~J., Engel, S.~M., Buckley, J.~P., and Edwards, J.~K.
  (2018).
\newblock A {{Bayesian}} approach to the g-formula.
\newblock {\em Statistical Methods in Medical Research}, 27(10):3183--3204.

\bibitem[Kukull et~al., 2002]{kukull2002dementia}
Kukull, W.~A., Higdon, R., Bowen, J.~D., McCormick, W.~C., Teri, L.,
  Schellenberg, G.~D., {van Belle}, G., Jolley, L., and Larson, E.~B. (2002).
\newblock Dementia and {{Alzheimer}} disease incidence: A prospective cohort
  study.
\newblock {\em Archives of Neurology}, 59(11):1737.

\bibitem[Lee et~al., 2017]{lee2017accelerated}
Lee, K.~H., Rondeau, V., and Haneuse, S. (2017).
\newblock Accelerated failure time models for semi-competing risks data in the
  presence of complex censoring.
\newblock {\em Biometrics}, 73(4):1401--1412.

\bibitem[Pang et~al., 2021]{pang2021flexible}
Pang, M., Platt, R.~W., Schuster, T., and Abrahamowicz, M. (2021).
\newblock Flexible extension of the accelerated failure time model to account
  for nonlinear and time-dependent effects of covariates on the hazard.
\newblock {\em Statistical Methods in Medical Research}, 30(11):2526--2542.

\bibitem[Portnoy, 2003]{portnoy2003censored}
Portnoy, S. (2003).
\newblock Censored regression quantiles.
\newblock {\em Journal of the American Statistical Association},
  98(464):1001--1012.

\bibitem[Prentice and Kalbfleisch, 1979]{prentice1979hazard}
Prentice, R.~L. and Kalbfleisch, J.~D. (1979).
\newblock Hazard rate models with covariates.
\newblock {\em Biometrics}, 35(1):25.

\bibitem[Reich and Smith, 2013]{reich2013bayesian}
Reich, B.~J. and Smith, L.~B. (2013).
\newblock Bayesian quantile regression for censored data.
\newblock {\em Biometrics}, 69(3):651--660.

\bibitem[Rothman et~al., 2021]{rothman2021modern}
Rothman, K.~J., Lash, T.~L., VanderWeele, T.~J., and Haneuse, S. (2021).
\newblock {\em Modern Epidemiology}.
\newblock {Wolters Kluwer}, {Philadelphia}, fourth edition edition.

\bibitem[Royston and Parmar, 2002]{royston2002flexible}
Royston, P. and Parmar, M. K.~B. (2002).
\newblock Flexible parametric proportional-hazards and proportional-odds models
  for censored survival data, with application to prognostic modelling and
  estimation of treatment effects.
\newblock {\em Statistics in Medicine}, 21(15):2175--2197.

\bibitem[Sj{\"o}lander, 2016]{sjolander2016regression}
Sj{\"o}lander, A. (2016).
\newblock Regression standardization with the {{R}} package {{stdReg}}.
\newblock {\em European Journal of Epidemiology}, 31(6):563--574.

\bibitem[{Stan Development Team}, 2020]{standevelopmentteam2020rstan}
{Stan Development Team} (2020).
\newblock {{RStan}}: The {{R}} interface to {{Stan}}.

\bibitem[Uno et~al., 2015]{uno2015alternatives}
Uno, H., Wittes, J., Fu, H., Solomon, S.~D., Claggett, B., Tian, L., Cai, T.,
  Pfeffer, M.~A., Evans, S.~R., and Wei, L.-J. (2015).
\newblock Alternatives to hazard ratios for comparing the efficacy or safety of
  therapies in noninferiority studies.
\newblock {\em Annals of Internal Medicine}, 163(2):127--134.

\bibitem[Vehtari et~al., 2017]{vehtari2017practical}
Vehtari, A., Gelman, A., and Gabry, J. (2017).
\newblock Practical {{Bayesian}} model evaluation using leave-one-out
  cross-validation and {{WAIC}}.
\newblock {\em Statistics and Computing}, 27(5):1413--1432.

\bibitem[Verdinelli and Wasserman, 1995]{verdinelli1995computing}
Verdinelli, I. and Wasserman, L. (1995).
\newblock Computing {{Bayes}} factors using a generalization of the
  {{Savage-Dickey}} density ratio.
\newblock {\em Journal of the American Statistical Association},
  90(430):614--618.

\bibitem[Wei, 1992]{wei1992accelerated}
Wei, L.~J. (1992).
\newblock The accelerated failure time model: {{A}} useful alternative to the
  cox regression model in survival analysis.
\newblock {\em Statistics in Medicine}, 11(14-15):1871--1879.

\bibitem[Zhang et~al., 2019]{zhang2019bayes}
Zhang, J., Hanson, T., and Zhou, H. (2019).
\newblock Bayes factors for choosing among six common survival models.
\newblock {\em Lifetime Data Analysis}, 25(2):361--379.

\bibitem[Zhou and Hanson, 2018]{zhou2018unified}
Zhou, H. and Hanson, T. (2018).
\newblock A unified framework for fitting bayesian semiparametric models to
  arbitrarily censored survival data, including spatially referenced data.
\newblock {\em Journal of the American Statistical Association},
  113(522):571--581.

\bibitem[Zhou et~al., 2017]{zhou2017generalized}
Zhou, H., Hanson, T., and Zhang, J. (2017).
\newblock Generalized accelerated failure time spatial frailty model for
  arbitrarily censored data.
\newblock {\em Lifetime Data Analysis}, 23(3):495--515.

\bibitem[Zhou et~al., 2020]{zhou2020spbayessurv}
Zhou, H., Hanson, T., and Zhang, J. (2020).
\newblock {{spBayesSurv}}: Fitting bayesian spatial survival models using
  {{R}}.
\newblock {\em Journal of Statistical Software}, 92(9).

\end{thebibliography}

\newpage

\appendix 

\makeatletter
\renewcommand{\thetable}{\thesection.\@arabic\c@table}
\@addtoreset{table}{section}
\makeatother

\makeatletter
\renewcommand{\thefigure}{\thesection.\@arabic\c@figure}
\@addtoreset{figure}{section}
\makeatother

\makeatletter
\renewcommand{\theequation}{\thesection.\@arabic\c@equation}
\@addtoreset{equation}{section}
\makeatother

\makeatletter
\renewcommand{\thelemma}{\thesection.\@arabic\c@lemma}
\@addtoreset{lemma}{section}
\makeatother


\section*{Appendix Introduction} \label{sec:appintro}

In these supplementary materials we present additional details and results beyond what could be presented in the main manuscript. To distinguish the two documents, alpha-numeric labels are used in this document while numeric labels are used in the main paper. Appendix~\ref{sec:addldata} provides additional results from the data application. Appendix~\ref{sec:appsims} presents results from a simulation study. Appendix~\ref{sec:vinvderiv} provides derivation of the form of $V^{-1}$ when $V$ is specified as a piecewise linear function of time. Appendix~\ref{sec:tvcovderiv} provides derivation of the form of the acceleration factor associated with a binary time-varying covariate. Appendix~\ref{sec:bayesfactor} derives an approximate Bayes Factor for comparison of accelerated failure time models with and without flexible effect specification. Appendix~\ref{sec:tbp_detail} provides additional detail on the transformed Bernstein polynomial (TBP) prior specification.

\section{Additional Data Application Results}\label{sec:addldata}
\subsection{AD/Dementia Onset}

In this section we report additional regression-standardized survival curves and acceleration factors for the onset of AD or dementia by APOE-$\epsilon4$ genetic variant status, for alternative specifications of the baseline distribution. We note that the most substantial difference between specifications occurs in the lowest quantiles, which represent parametric extrapolation beyond the observed data quantiles.

\begin{figure}[H]

\begin{center}
	\minipage{0.5\textwidth}%
		\includegraphics[width=\linewidth]{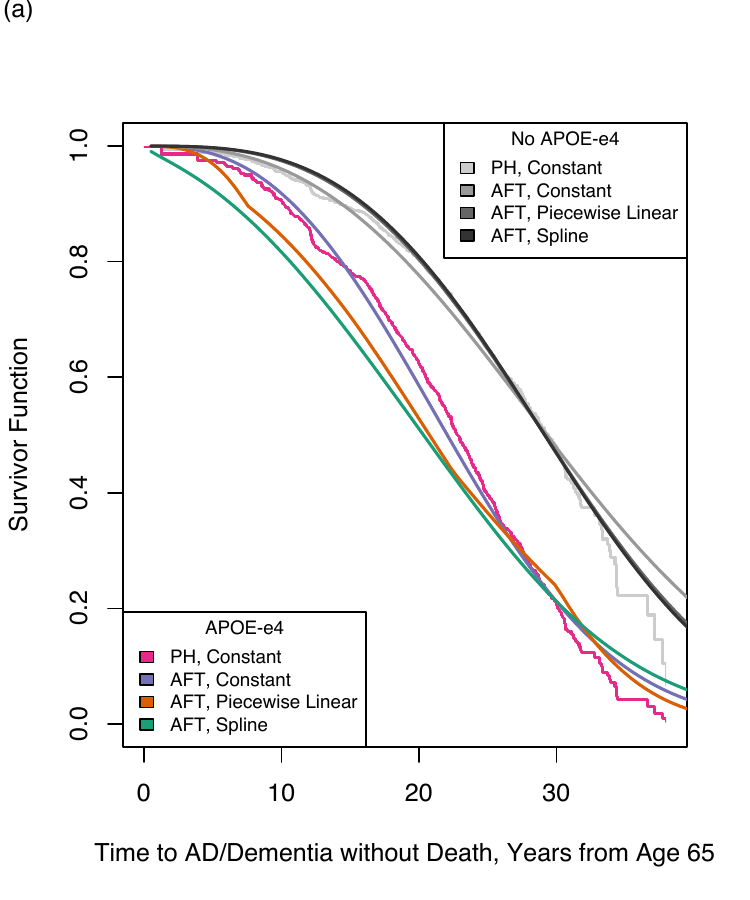}		\endminipage\hfill
	\minipage{0.5\textwidth}
	\includegraphics[width=\linewidth]{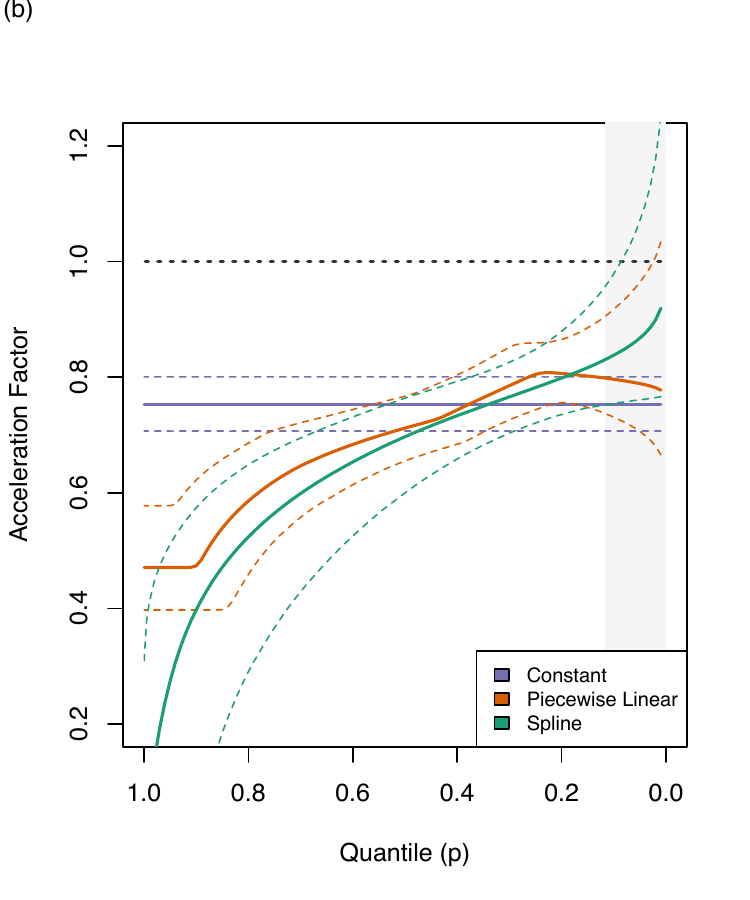}
	\endminipage
\end{center}
\caption{Under a Weibull baseline specification: (a) regression standardized survivor function estimates for onset of AD or dementia without death, averaged over other covariates. Regression standardized estimate from Cox proportional hazards model shown for comparison; (b) regression standardized quantile-varying acceleration factor estimates for onset of AD or dementia without death, averaged over other covariates. 95\% credible intervals represented with dashed lines. Grey shaded region represents area of parametric extrapolation beyond quantiles observed in both groups. \label{fig:ad_marg_surv_af_wb}}
\end{figure}

\begin{figure}[H]
\begin{center}
	\minipage{0.5\textwidth}%
		\includegraphics[width=\linewidth]{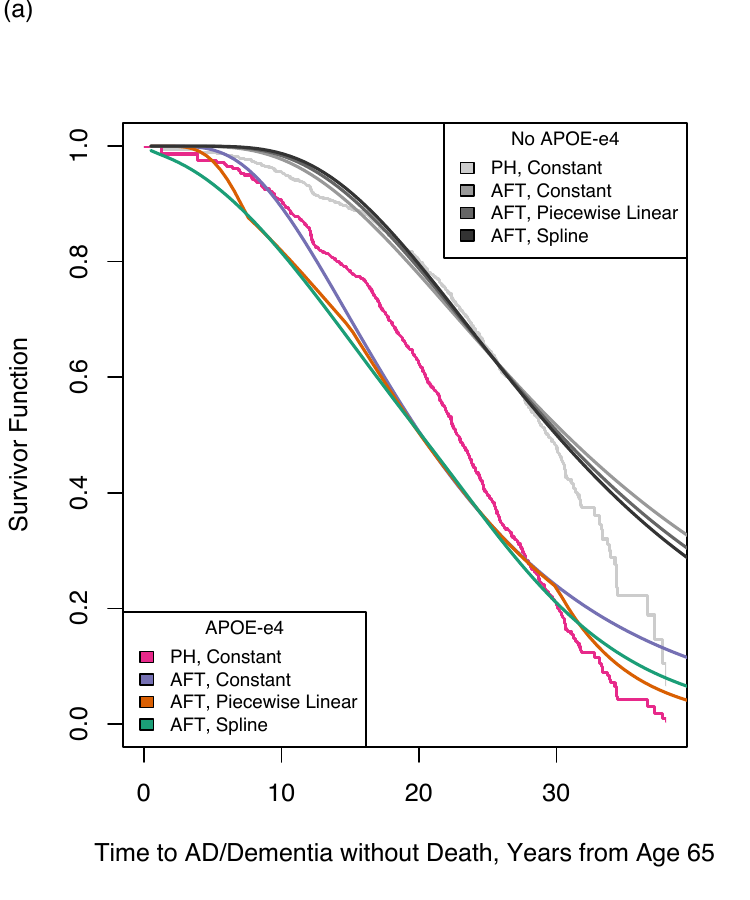}		\endminipage\hfill
	\minipage{0.5\textwidth}
	\includegraphics[width=\linewidth]{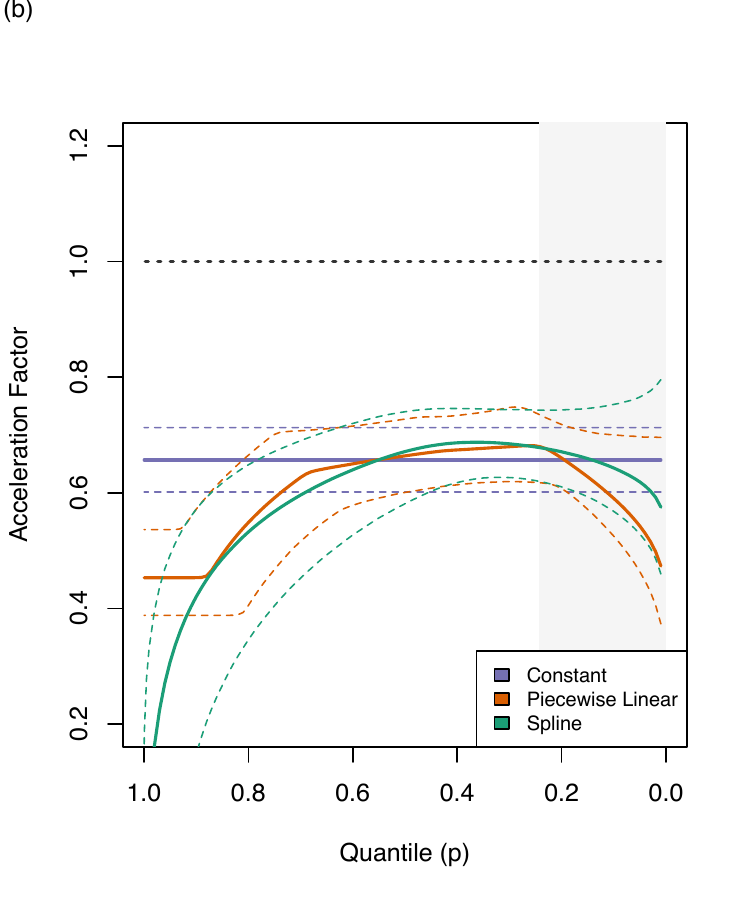}
	\endminipage
\end{center}
\caption{Under a log-Normal baseline specification: (a) regression standardized survivor function estimates for onset of AD or dementia without death, averaged over other covariates. Regression standardized estimate from Cox proportional hazards model shown for comparison; (b) regression standardized quantile-varying acceleration factor estimates for onset of AD or dementia without death, averaged over other covariates. 95\% credible intervals represented with dashed lines. Grey shaded region represents area of parametric extrapolation beyond quantiles observed in both groups.  \label{fig:ad_marg_surv_af_ln}}
\end{figure}

\subsection{Mortality following AD/Dementia Onset}

Below we report regression parameter estimates, and additional regression-standardized survivor curves and acceleration factors for mortality by AD/dementia status, across alternative specifications for the baseline distribution.

\begin{figure}[H]

\begin{center}
	\minipage{0.48\textwidth}%
		\includegraphics[width=\linewidth]{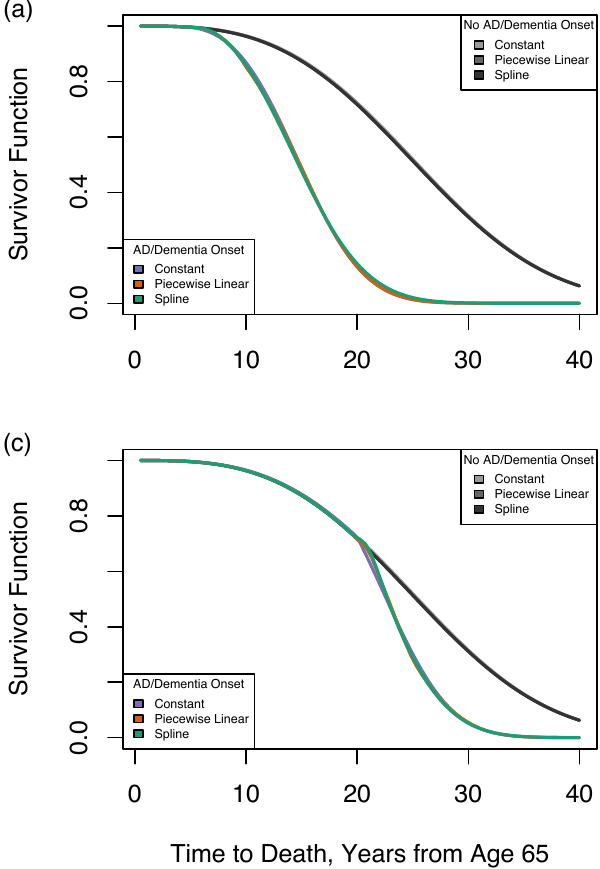}		\endminipage\hfill
	\minipage{0.48\textwidth}
	\includegraphics[width=\linewidth]{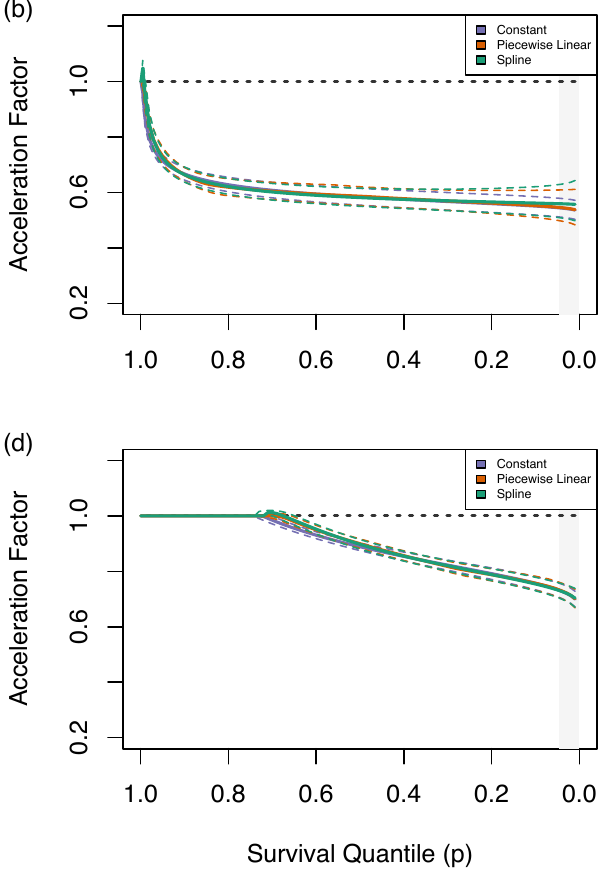}
	\endminipage
\end{center}
\caption{Under a Weibull baseline specification: regression standardized survivor function estimates for mortality following onset of AD or dementia at (a) age 70 and (c) age 85, averaged over other covariates; regression standardized survivor function estimates for mortality following onset of AD or dementia at (b) age 70 and (d) age 85, averaged over other covariates. 95\% credible intervals represented with dashed lines. Grey shaded region represents area of parametric extrapolation beyond quantiles observed in both groups.  \label{fig:deathtv_marg_surv_af_wb}}
\end{figure}

\begin{figure}[H]
\begin{center}
	\minipage{0.48\textwidth}%
		\includegraphics[width=\linewidth]{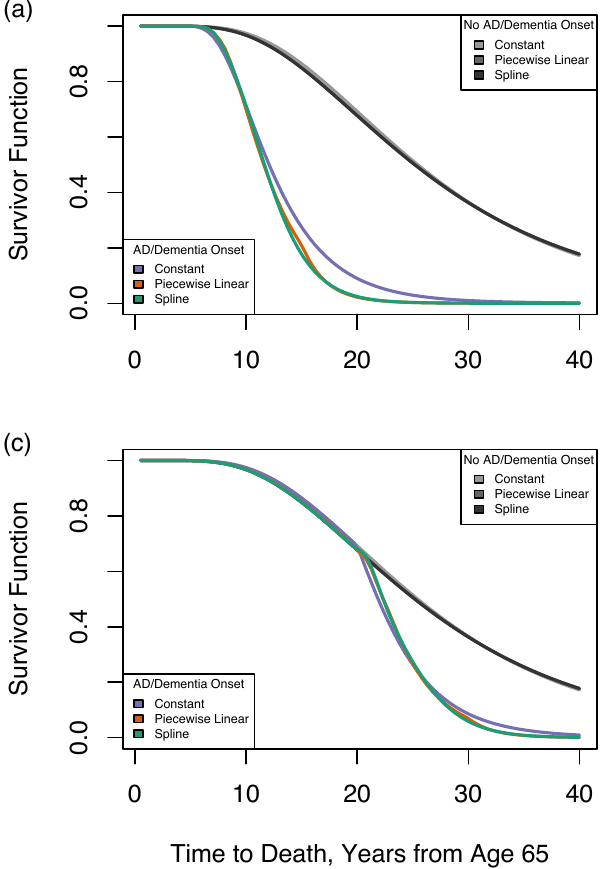}		\endminipage\hfill
	\minipage{0.48\textwidth}
	\includegraphics[width=\linewidth]{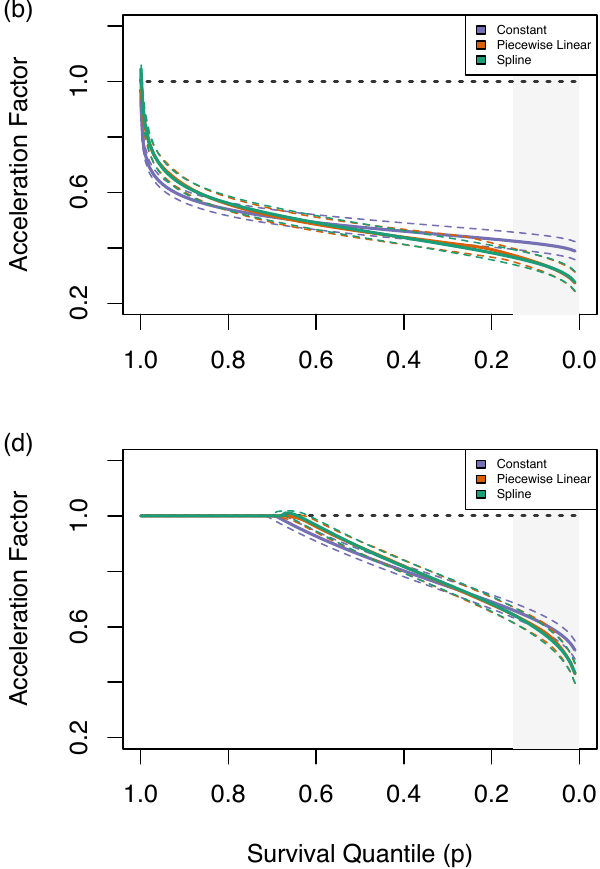}
	\endminipage
\end{center}
\caption{Under a log-Normal baseline specification: regression standardized survivor function estimates for mortality following onset of AD or dementia at (a) age 70 and (c) age 85, averaged over other covariates; regression standardized survivor function estimates for mortality following onset of AD or dementia at (b) age 70 and (d) age 85, averaged over other covariates. 95\% credible intervals represented with dashed lines. Grey shaded region represents area of parametric extrapolation beyond quantiles observed in both groups.  \label{fig:deathtv_marg_surv_af_ln}}
\end{figure}

\begin{figure}[H]
\begin{center}
	\minipage{0.33\textwidth}%
		\includegraphics[width=\linewidth]{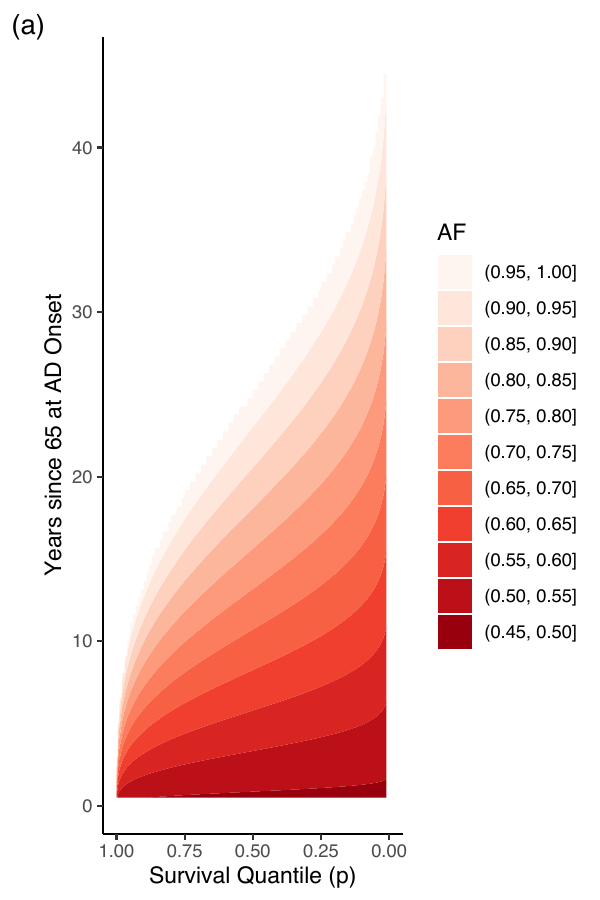}		\endminipage\hfill
	\minipage{0.33\textwidth}
	\includegraphics[width=\linewidth]{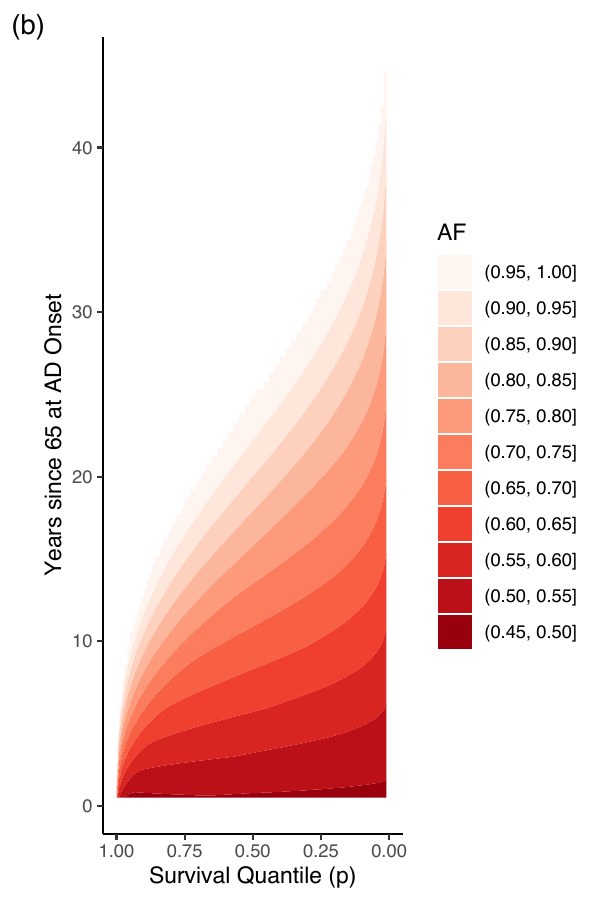}		\endminipage\hfill
	\minipage{0.33\textwidth}
	\includegraphics[width=\linewidth]{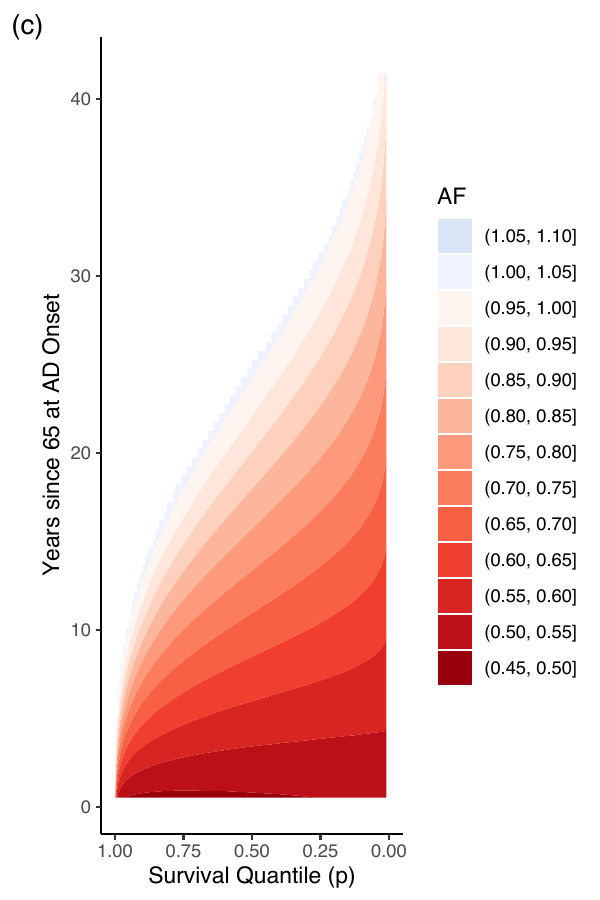}	
	\endminipage
\end{center}
\caption{Under a Weibull baseline specification, contour plots of regression standardized acceleration factor surface estimates for death following onset of AD/dementia, averaged over other covariates. Time of AD/dementia onset is shown on y-axis, and subsequent survival quantile is shown on x-axis. Color indicates acceleration factor at the given survival quantile, comparing those with AD/dementia onset at the specified time and those without AD/dementia. Horizontal cross-sections illustrate quantile-varying acceleration factor for AD/dementia onset at a particular time, while vertical cross-sections illustrate acceleration factor at a particular quantile across times of AD/dementia onset. (a) constant effect specification; (b) piecewise linear effect specification; and (c) spline effect specification. \label{fig:deathtv_marg_afsurface_wb}}
\end{figure}

\begin{figure}[H]
\begin{center}
	\minipage{0.33\textwidth}%
		\includegraphics[width=\linewidth]{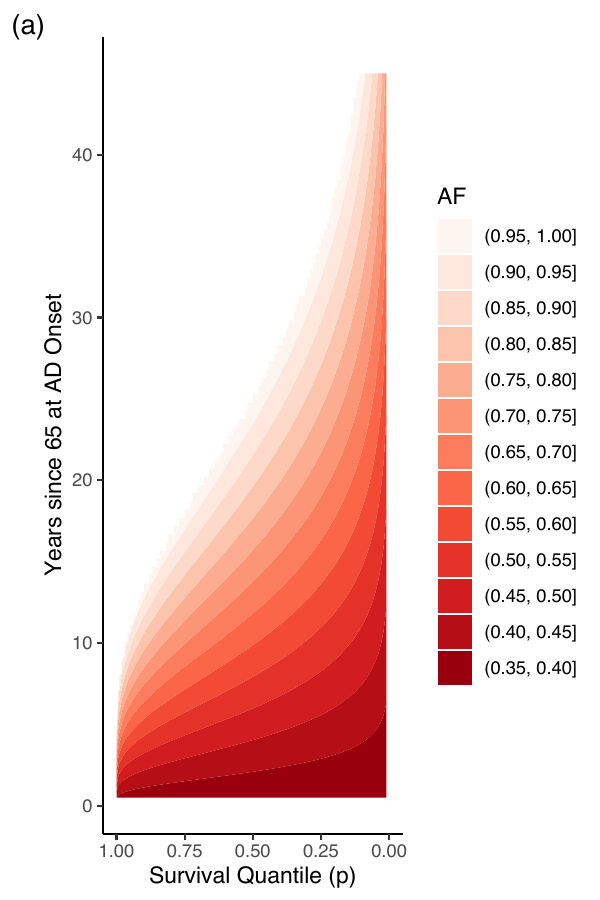}		\endminipage\hfill
	\minipage{0.33\textwidth}
	\includegraphics[width=\linewidth]{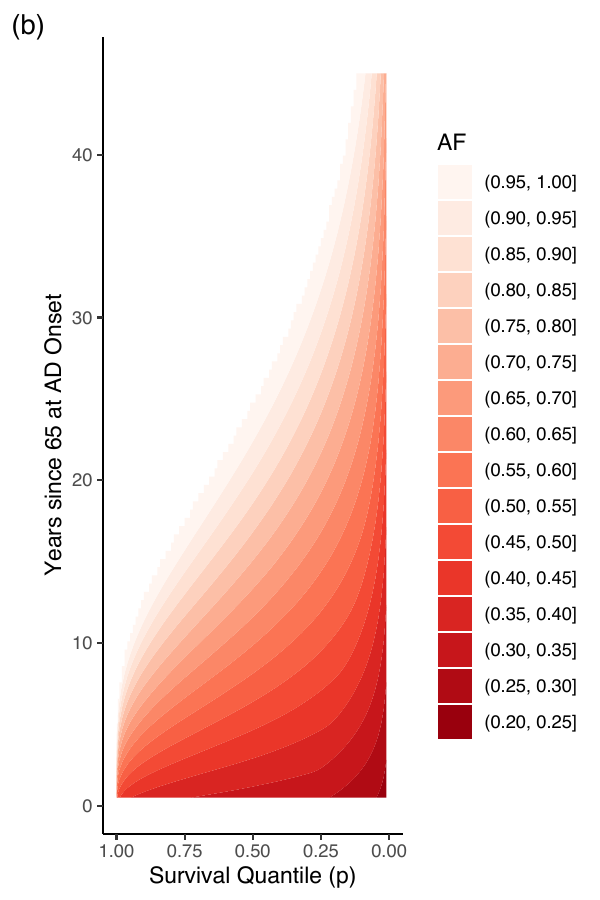}		\endminipage\hfill
	\minipage{0.33\textwidth}
	\includegraphics[width=\linewidth]{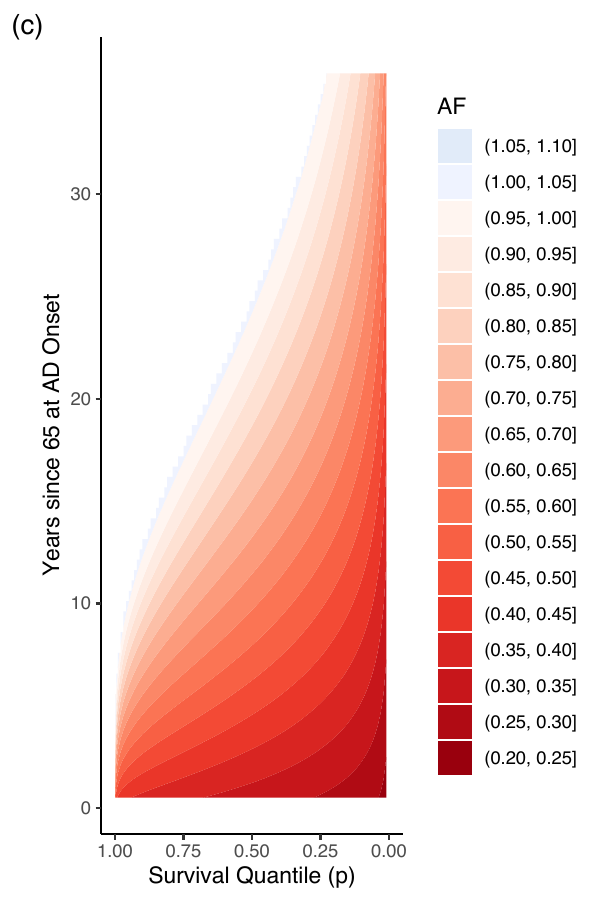}	
	\endminipage
\end{center}
\caption{Under a log-Normal baseline specification, contour plots of regression standardized acceleration factor surface estimates for death following onset of AD/dementia, averaged over other covariates. Time of AD/dementia onset is shown on y-axis, and subsequent survival quantile is shown on x-axis. Color indicates acceleration factor at the given survival quantile, comparing those with AD/dementia onset at the specified time and those without AD/dementia. Horizontal cross-sections illustrate quantile-varying acceleration factor for AD/dementia onset at a particular time, while vertical cross-sections illustrate acceleration factor at a particular quantile across times of AD/dementia onset. (a) constant effect specification; (b) piecewise linear effect specification; and (c) spline effect specification. \label{fig:deathtv_marg_afsurface_ln}}
\end{figure}

\begin{table}[H]
\begin{center}
\caption{Regression estimates for time to death. AFT results are posterior medians and 95\% credible intervals for regression parameters. Cox model results are log-hazard ratio estimates and 95\% confidence intervals. \label{tab:deathtv_results}}
\resizebox{0.95\textwidth}{!}{\begin{tabular}{lcccc}
  \hline
& & \multicolumn{3}{c}{AFT Model} \\ \cline{3-5}
 & Cox PH & log-Normal & Weibull & TBP (Weibull Centered) \\ 
  \hline
White Race/Ethnicity, $\beta_1$ &  &  &  &  \\ 
~~Constant & 0.17 (-0.07, 0.41) & -0.07 (-0.17, 0.03) & -0.08 (-0.16, 0) & -0.07 (-0.15, 0) \\ 
~~Piecewise Linear &  & -0.1 (-0.21, 0.01) & -0.08 (-0.16, 0) & -0.07 (-0.15, 0) \\ 
~~Restricted Cubic Spline &  & -0.1 (-0.21, 0) & -0.08 (-0.16, 0) & -0.07 (-0.15, 0) \\ 
  Male Sex, $\beta_2$ &  &  &  &  \\ 
~~Constant & 0.52 (0.41, 0.64) & -0.25 (-0.31, -0.2) & -0.16 (-0.2, -0.12) & -0.14 (-0.17, -0.11) \\ 
~~Piecewise Linear &  & -0.27 (-0.33, -0.21) & -0.16 (-0.2, -0.13) & -0.14 (-0.18, -0.11) \\ 
~~Restricted Cubic Spline &  & -0.27 (-0.33, -0.21) & -0.16 (-0.2, -0.13) & -0.14 (-0.18, -0.11) \\ 
  Married at Study Entry, $\beta_3$ &  &  &  &  \\ 
~~Constant & -0.16 (-0.3, -0.01) & 0.12 (0.05, 0.19) & 0.05 (0.01, 0.1) & 0.04 (0, 0.08) \\ 
~~Piecewise Linear &  & 0.12 (0.05, 0.19) & 0.05 (0.01, 0.1) & 0.04 (0, 0.08) \\ 
~~Restricted Cubic Spline &  & 0.12 (0.05, 0.19) & 0.05 (0.01, 0.1) & 0.04 (0, 0.08) \\ 
  $\geq$15 Years of Education, $\beta_4$ &  &  &  &  \\ 
~~Constant & -0.1 (-0.21, 0.01) & 0.07 (0.02, 0.13) & 0.03 (-0.01, 0.06) & 0.02 (-0.01, 0.05) \\ 
~~Piecewise Linear &  & 0.09 (0.03, 0.15) & 0.03 (-0.01, 0.07) & 0.02 (-0.01, 0.05) \\ 
~~Restricted Cubic Spline &  & 0.09 (0.03, 0.15) & 0.03 (0, 0.07) & 0.02 (-0.01, 0.05) \\ 
  APOE-$\epsilon4$ Genetic Variant, $\beta_5$ &  &  &  &  \\ 
~~Constant & 0.01 (-0.11, 0.13) & 0.06 (0, 0.11) & 0.01 (-0.03, 0.05) & 0 (-0.03, 0.04) \\ 
~~Piecewise Linear &  & 0.06 (0, 0.12) & 0.01 (-0.03, 0.05) & 0 (-0.03, 0.04) \\ 
~~Restricted Cubic Spline &  & 0.06 (0, 0.12) & 0.01 (-0.03, 0.05) & 0 (-0.03, 0.04) \\ 
AD/Dementia Onset, $\beta_6$ &  &  &  &  \\ 
~~Constant & 1.14 (1.02, 1.26) & -1.06 (-1.16, -0.97) & -0.73 (-0.81, -0.65) & -0.67 (-0.75, -0.6) \\ 
~~Piecewise Linear &  & -0.14 (-0.42, 0.17) & -0.02 (-0.3, 0.29) & 0.01 (-0.27, 0.32) \\ 
~~Restricted Cubic Spline &  & 1.59  (0.90, 2.35) & 1.63 (0.95, 2.38) & 1.68  (1.00, 2.42) \\ 
AD/Dementia Onset, $\alpha_1$ &  &  &  &  \\ 
~~Constant &  &  &  &  \\ 
~~Piecewise Linear &  & -0.95 (-1.29, -0.63) & -0.86 (-1.21, -0.53) & -0.84 (-1.19, -0.52) \\ 
~~Restricted Cubic Spline &  & -1.77 (-2.25, -1.33) & -1.60 (-2.08, -1.17) & -1.57 (-2.03, -1.13) \\ 
AD/Dementia Onset, $\alpha_2$ &  &  &  &  \\ 
~~Constant &  &  &  &  \\ 
~~Piecewise Linear &  & -1.15 (-1.49, -0.82) & -0.90 (-1.25, -0.57) & -0.86 (-1.2, -0.53) \\ 
~~Restricted Cubic Spline &  & -5.07 (-6.56, -3.73) & -4.40 (-5.86, -3.10) & -4.42 (-5.85, -3.10) \\ 
AD/Dementia Onset, $\alpha_3$ &  &  &  &  \\ 
~~Constant &  &  &  &  \\ 
~~Piecewise Linear &  & -1.15 (-1.49, -0.82) & -0.65 (-0.99, -0.33) & -0.61 (-0.96, -0.29) \\ 
~~Restricted Cubic Spline &  & -1.78 (-2.18, -1.41) & -1.08 (-1.45, -0.75) & -1.07 (-1.43, -0.73) \\ 
AD/Dementia Onset, $\alpha_4$ &  &  &  &  \\ 
~~Constant &  &  &  &  \\ 
~~Piecewise Linear &  & -1.68 (-2.11, -1.26) & -0.71 (-1.10, -0.32) & -0.73 (-1.15, -0.31) \\ 
~~Restricted Cubic Spline &  &  &  &  \\ 
   \hline
\end{tabular}}
\end{center}
\end{table}

\subsection{Additional Comparisons with other Bayesian Survival Models} \label{sec:tbp_detail}

The \texttt{spBayesSurv} package implements the TBP prior for other survival effect specifications beyond the accelerated failure time model, including proportional hazards and proportional odds models, using an adaptive Metropolis Hastings sampler \citep{zhou2020spbayessurv}. Table~\ref{tab:diagnosticsspb} presents the model fit metrics for comparator proportional hazards and proportional odds models from this package, while Table~\ref{tab:spbresults} presents the estimated coefficients of each covariate. Baseline prior details are the same as the TBP models reported in the main text.

As expected, different effect specifications are suited to different application settings, and conceptually we reiterate the appealing interpretation of accelerated failure time models in the study of AD and dementia in older adults. Nonetheless, we do see that model fit metrics are comparable to the AFT models presented in the main text, and that qualitatively the estimated effects are concordant.

\begin{table}[H]
\centering
\caption{Estimated log-pseudo marginal likelihood (LPML), multiplied by -2 to replicate scale of information criteria. Smaller values indicate better model fit, and are comparable to ELPD metric.
\label{tab:diagnosticsspb}}
\resizebox{\textwidth}{!}{\begin{tabular}{lcc}
  \hline
& \multicolumn{2}{c}{TBP (Weibull Centered) } \\ \cline{2-3}
 & Proportional Hazards & Proportional Odds \\ 
  \hline
AD/Dementia Onset (Death as a Censoring Mechanism) & 5793.2 & 5775.5 \\   
Death (AD/Dementia as a Time-Varying Covariate) & 9589.5 & 9638.3 \\   
\hline
\end{tabular}}
\end{table}

\begin{table}[H]
\begin{center}
\caption{Additional regression coefficient posterior means and 95\% credible intervals. \label{tab:spbresults}}
\begin{tabular}{lcc}
  \hline
& \multicolumn{2}{c}{TBP (Weibull Centered) } \\ \cline{2-3}
 & Proportional Hazards & Proportional Odds \\ 
\hline
\multicolumn{3}{l}{\textbf{AD/Dementia Onset (Death as a Censoring Mechanism)}} \\   
\hline
~~White Race/Ethnicity & -0.24 (-0.53, 0.07) & -0.45 (-0.88, -0.05) \\ 
~~Male Sex & 0.06 (-0.13, 0.23) & 0.13 (-0.11, 0.38) \\ 
~~Married at Study Entry & -0.27 (-0.49, -0.06) & -0.41 (-0.72, -0.13) \\ 
~~$\geq$15 Years of Education & -0.11 (-0.26, 0.05) & -0.12 (-0.35, 0.11) \\ 
~~APOE-$\epsilon4$ Genetic Variant & 0.76 (0.60, 0.92) & 1.22 (0.99, 1.46) \\ 
\hline
\multicolumn{3}{l}{\textbf{Death (AD/Dementia as a Time-Varying Covariate)}} \\   
\hline
~~White Race/Ethnicity & 0.18 (-0.05, 0.43) & 0.05 (-0.39, 0.50) \\ 
~~Male Sex & 0.52 (0.40, 0.64) & 0.88 (0.66, 1.11) \\ 
~~Married at Study Entry & -0.17 (-0.32, -0.01) & -0.42 (-0.69, -0.17) \\ 
~~$\geq$15 Years of Education & -0.10 (-0.20, 0.00) & -0.18 (-0.40, 0.04) \\ 
~~APOE-$\epsilon4$ Genetic Variant & 0.01 (-0.11, 0.13) & -0.09 (-0.33, 0.15) \\ 
~~AD/Dementia Onset & 1.14 (1.03, 1.25) & 2.92 (2.40, 3.62) \\
\hline
\end{tabular}
\end{center}
\end{table}

\section{Simulation Results} \label{sec:appsims}

To examine the operating characteristics of the proposed methodology, we performed a simulation study presented here. The purpose of this simulation study is threefold: (1) validate the proposed method in recovering an underlying data generating mechanism, (2) assess estimation performance of the acceleration factor across quantiles in the presence of right censoring, and (3) assess performance of marginal acceleration factor estimation via regression standardization.

For these simulations we considered a binary exposure $X_1\sim\text{Binomial}(0.5)$ with potentially quantile-varying effect, and two additional continuous covariates $X_2$ and $X_3$ each drawn from a standard normal. We specified a true baseline distribution using parameters comparable to the Alzheimer's disease data under study, as a log-Normal baseline hazard with $\mu = \log(3.2)$ and $\sigma = 0.55$ and two covariate effect scenarios: a true constant acceleration factor, and a true piecewise quantile-varying acceleration factor with 4 breakpoints at 7.5 year intervals between 0 and 30, each again structured after the observed results from the data application. Under a model of the form in Equation (3.13) the corresponding true parameters for each scenario are provided in Table \ref{tab:simsettings}. We further generate random right censoring uniformly between 15 and 40, corresponding to the age interval 80-105 in the study based on a time origin of age 65. The resulting overall censoring rate was similar in each scenario, about 45\%.

\begin{table}[H]
\begin{center}
\caption{True data generating mechanisms for simulation study. \label{tab:simsettings}}
\resizebox{\textwidth}{!}{\begin{tabular}{lccccc}
  \hline
Scenario & $\bbeta$ & $\balpha$ & \multicolumn{3}{c}{True $X_1$ Acceleration Factor $(p = 0.75, 0.5, 0.25)$} \\ \cline{4-6}%
 &  &  & Conditional on $X_2=X_3=0$ && Marginal over $X_2,X_3\sim N(0,1)$ \\ \hline
Constant Effect& $(-0.2, -0.5, 0.5)\trans$ & $(0,0,0,0)\trans$  & (0.818, 0.818, 0.818) && (0.818, 0.818, 0.818) \\
Piecewise Effect & $(-0.2, -0.5, 0.5)\trans$ & $(0, 0.3,  0.45, 0.5)\trans$ & (0.818, 0.891, 1.018) && (0.818, 0.891, 1.087) \\
   \hline
\end{tabular}}
\end{center}
\end{table}

For each data generating mechanism we generated $n=2000$ observations, and for each model we used the same prior specifications as the main text, and drew 4000 posterior samples with 50\% adaptation/burn-in from a single chain. We calculated the following performance metrics over $300$ simulations:
\begin{itemize}
	\item median ELPD model metric,
	\item median integrated squared difference in the estimated and true hazard, $\int[h_0(t)-h_0^{*}(t)]^2dt$ over the interval $[0,40]$,
	\item median bias, standard deviation, and 95\%CrI coverage probability of the estimated acceleration factor for $X_1$, evaluated at the 0.75, 0.5 and 0.25 quantiles both conditionally on $X_2=X_3=0$ regression standardized over the empirical distributions of $X_2$ and $X_3$. The `true' marginal acceleration factors used for reference were computed using Gauss-Hermite numerical quadrature over the standard normal nuisance covariates.
\end{itemize}

This was purposefully a challenging set of simulations, with high rates of censoring in the lower quantiles as well as an attenuating effect that switched from harmful in the higher quantiles to protective in the lower quantiles. Therefore, it was expected that performance would degrade particularly by the 0.25 quantile, as in many simulated datasets there is little to no available data in this part of the distribution.

Despite this, the simulations did not exhibit any computational challenges estimating the fully parametric models. However, in simulations evaluating the TBP prior model, we found that the Stan implementation exhibited some fragility depending on the dataset and algorithm initialization. In a minority of cases, the adaptive No-U-Turn sampler implemented in Stan became stuck at the initial value and selected an infinitesimal sampler stepsize, yielding an algorithm that did not explore the posterior at all. We believe this is due to the algorithm's use of posterior gradient information calculated numerically using finite differences, which have potential for underflow of the TBP summands despite implementation using computational techniques to improve stability. (E.g., stable implementations of $\log(1+x)$ and $\exp(x)-1$ in Stan, computation on the log scale using the logSumExp trick, and following the Adaptive Monte Carlo implementation of this baseline hazard available in the \texttt{spBayesSurv} package, thresholding for summands smaller than machine precision).

Across all settings and effect specifications, this occurred between 20-35\% of the time on first initialization, and therefore the below results for TBP prior models are restricted to those that successfully sampled with all split-chain $\Rhat$ (or `potential scale reduction factor') values below 1.05 and at least 100 effective samples for every parameter. In practical applications, these issues may be addressed through additional tuning of choice of initial values or modification of the priors which is not practical in a simulation context, but because this is to our knowledge the first implementation of a survival model with TBP prior baseline using the Stan language and No-U-Turn Hamiltonian Monte Carlo sampling algorithm, we believe that these observations provide valuable insight into the properties and potential wider use of this model specification.

\subsection{Baseline Hazard Estimation and Overall Model Performance Metrics}

Summarized in Table \ref{tab:modelmetric}, as measured by the mean integrated squared difference of the baseline hazard the correctly specified parametric log-Normal model best recovered the true baseline hazard, followed by the TBP prior model, while the Weibull parametric model had the largest error.

On average, models with correctly specified effects as well as correctly specified baselines had the best ELPD. However, under the truly quantile-varying data generating mechanism, the log-Normal model with constant acceleration factor underperformed models with TBP-prior baselines and some form of quantile-varying effect. Finally, we saw that models with misspecified Weibull baseline uniformly had underperforming metrics relative to other models, though Weibull models with quantile-varying effects outperformed the standard Weibull AFT.

\begin{table}[ht]
\centering
\caption{Median ELPD and median integrated squared error of the baseline hazard.  \label{tab:modelmetric}}
\begin{tabular}{lcc}
  \hline
Simulation Setting & $-2*\log(\widehat{\text{ELPD}}_{\text{psis-loo}})$ & $\int_0^{40} (\widehat{h}_0(t) - h^*_0(t))^2dt$ \\ 
  \hline
  \multicolumn{3}{l}{\textbf{Constant True Effect}} \\
   \multicolumn{3}{l}{~~\textit{Constant Effect Specification}} \\
   ~~~~Log-Normal & 8489.6 & 0.000097 \\ 
   ~~~~TBP & 8507.1 & 0.000640 \\ 
   ~~~~Weibull & 8646.7 & 0.004841 \\ 
   \multicolumn{3}{l}{~~\textit{Piecewise Effect Specification}} \\
  ~~~~Log-Normal & 8492.8 & 0.000138 \\ 
  ~~~~TBP & 8506.3 & 0.000794 \\ 
  ~~~~Weibull & 8645.6 & 0.007231 \\ 
  \multicolumn{3}{l}{~~\textit{Spline Effect Specification}} \\
  ~~~~Log-Normal & 8490.6 & 0.000146 \\ 
  ~~~~TBP & 8510.0 & 0.000712 \\ 
  ~~~~Weibull & 8637.5 & 0.005453 \\ 
  \multicolumn{3}{l}{\textbf{Piecewise True Effect}} \\
  \multicolumn{3}{l}{~~\textit{Constant Effect Specification}} \\
  ~~~~Log-Normal & 8201.6 & 0.000721 \\ 
  ~~~~TBP & 8218.9 & 0.001521 \\ 
  ~~~~Weibull & 8379.2 & 0.003102 \\ 
  \multicolumn{3}{l}{~~\textit{Piecewise Effect Specification}} \\
  ~~~~Log-Normal & 8184.9 & 0.000151 \\ 
  ~~~~TBP & 8187.1 & 0.000696 \\ 
  ~~~~Weibull & 8336.7 & 0.007370 \\ 
  \multicolumn{3}{l}{~~\textit{Spline Effect Specification}} \\
  ~~~~Log-Normal & 8185.6 & 0.000154 \\ 
  ~~~~TBP & 8203.2 & 0.000633 \\
  ~~~~Weibull & 8332.7 & 0.005468 \\ 
    \hline  
\end{tabular}
\end{table}

\subsection{Acceleration factor estimation performance}

Operating characteristics of conditional acceleration factors are reported in Table~\ref{tab:condafsim}, and regression standardized acceleration factors in Table~\ref{tab:margafsim}. When the parametric baseline was correctly specified, the AF estimates performed very well across quantiles under both data generating mechanisms. When the effect was truly quantile-invariant, the log-Normal models with piecewise and spline effects successfully captured the effect across quantiles with little bias or efficiency loss. When the effect truly varied across quantiles, the log-Normal piecewise model successfully recovered the truth, and the log-Normal spline model approximated this effect with little bias or efficiency loss across quantiles. These results held both for the conditional AF estimates setting $X_2=X_3=0$, and the regression-standardized AF estimates averaged over these additional covariates.

However, the AF estimation performance of the models with flexible TBP-prior baseline was somewhat reduced relative to the correctly-specified parametric model, especially in the lower quantiles where information loss was highest. Under the quantile-invariant data generating mechanism, the TBP-prior models had slightly elevated bias and mild undercoverage of the credible intervals, particularly for the models with flexibly-specified effects. This also occurred under the quantile-varying data generating mechanism, and as expected the performance degraded most heavily for the 25th quantile AF estimates. Nevertheless, the Weibull-centered TBP-prior uniformly outperformed the model with a misspecified parametric Weibull baseline.

These results support our decision in practical applications to use the model diagnostic criterion to select the final model specification, as the model diagnostic criterion was generally concordant with AF estimation performance.

\begin{table}[H]
\centering
\caption{Simulation results for conditional acceleration factors for $X_1$, setting $X_2=X_3=0$. TBP baseline specification is Weibull-centered. \label{tab:condafsim}}
\resizebox{\textwidth}{!}{\begin{tabular}{lccccccccccc}
  \hline
& \multicolumn{3}{c}{Bias} & & \multicolumn{3}{c}{Standard Deviation} & & \multicolumn{3}{c}{Coverage Probability}\\
\cline{2-4} \cline{6-8} \cline{10-12}
Quantile: & $0.75$ & $0.5$ & $0.25$ & & $0.75$ & $0.5$ & $0.25$ & & $0.75$ & $0.5$ & $0.25$ \\ 
  \hline
  \multicolumn{3}{l}{\textbf{Constant True Effect}} \\
   \multicolumn{3}{l}{~~\textit{Constant Effect Specification}} \\
  ~~~~Log-Normal & 0.001 & 0.001 & 0.001 &  & 0.023 & 0.023 & 0.023 &  & 0.940 & 0.940 & 0.940 \\ 
  ~~~~TBP & 0.009 & 0.009 & 0.009 &  & 0.023 & 0.023 & 0.023 &  & 0.922 & 0.922 & 0.922 \\ 
  ~~~~Weibull & 0.011 & 0.011 & 0.011 &  & 0.023 & 0.023 & 0.023 &  & 0.873 & 0.873 & 0.873 \\ 
   \multicolumn{3}{l}{~~\textit{Piecewise Effect Specification}} \\
  ~~~~Log-Normal & 0.003 & 0.001 & 0.001 &  & 0.026 & 0.027 & 0.034 &  & 0.933 & 0.937 & 0.940 \\ 
  ~~~~TBP & 0.007 & 0.015 & 0.028 &  & 0.026 & 0.026 & 0.033 &  & 0.936 & 0.900 & 0.836 \\ 
  ~~~~Weibull & -0.017 & 0.009 & 0.046 &  & 0.025 & 0.026 & 0.033 &  & 0.863 & 0.907 & 0.690 \\ 
   \multicolumn{3}{l}{~~\textit{Spline Effect Specification}} \\
  ~~~~Log-Normal & 0.002 & 0.001 & -0.003 &  & 0.025 & 0.028 & 0.034 &  & 0.920 & 0.933 & 0.953 \\ 
  ~~~~TBP & 0.002 & 0.012 & 0.022 &  & 0.025 & 0.026 & 0.033 &  & 0.923 & 0.887 & 0.861 \\ 
  ~~~~Weibull & -0.021 & 0.005 & 0.037 &  & 0.025 & 0.026 & 0.033 &  & 0.840 & 0.923 & 0.740 \\ 
  \multicolumn{3}{l}{\textbf{Piecewise True Effect}} \\
   \multicolumn{3}{l}{~~\textit{Constant Effect Specification}} \\
  ~~~~Log-Normal & 0.049 & -0.023 & -0.151 &  & 0.026 & 0.026 & 0.026 &  & 0.517 & 0.810 & 0.000 \\ 
  ~~~~TBP & 0.058 & -0.015 & -0.142 &  & 0.026 & 0.026 & 0.026 &  & 0.386 & 0.878 & 0.000 \\ 
  ~~~~Weibull & 0.078 & 0.005 & -0.122 &  & 0.027 & 0.027 & 0.027 &  & 0.187 & 0.897 & 0.017 \\ 
   \multicolumn{3}{l}{~~\textit{Piecewise Effect Specification}} \\
  ~~~~Log-Normal & 0.003 & 0.002 & 0.013 &  & 0.026 & 0.036 & 0.067 &  & 0.930 & 0.937 & 0.947 \\ 
  ~~~~TBP & 0.003 & 0.012 & 0.043 &  & 0.026 & 0.035 & 0.069 &  & 0.921 & 0.908 & 0.882 \\ 
  ~~~~Weibull & -0.017 & 0.031 & 0.089 &  & 0.025 & 0.037 & 0.073 &  & 0.863 & 0.867 & 0.683 \\ 
   \multicolumn{3}{l}{~~\textit{Spline Effect Specification}} \\
  ~~~~Log-Normal & 0.012 & -0.002 & 0.014 &  & 0.027 & 0.035 & 0.067 &  & 0.927 & 0.943 & 0.950 \\ 
  ~~~~TBP& 0.011 & 0.006 & 0.050 &  & 0.027 & 0.034 & 0.068 &  & 0.904 & 0.947 & 0.885 \\ 
  ~~~~Weibull & -0.010 & 0.019 & 0.080 &  & 0.027 & 0.035 & 0.071 &  & 0.887 & 0.903 & 0.723 \\ 
   \hline
\end{tabular}}
\end{table}

\begin{table}[H]
\centering
\caption{Simulation results for marginal acceleration factors for $X_1$ averaged over empirical distribution of $X_2$ and $X_3$. TBP baseline specification is Weibull-centered. \label{tab:margafsim}}
\resizebox{\textwidth}{!}{\begin{tabular}{lccccccccccc}
  \hline
& \multicolumn{3}{c}{Bias} & & \multicolumn{3}{c}{Standard Deviation} & & \multicolumn{3}{c}{Coverage Probability}\\
\cline{2-4} \cline{6-8} \cline{10-12}
Quantile: & $0.75$ & $0.5$ & $0.25$ & & $0.75$ & $0.5$ & $0.25$ & & $0.75$ & $0.5$ & $0.25$ \\ 
  \hline
  \multicolumn{3}{l}{\textbf{Constant True Effect}} \\
   \multicolumn{3}{l}{~~\textit{Constant Effect Specification}} \\
  ~~~~Log-Normal & 0.001 & 0.001 & 0.001 &  & 0.023 & 0.023 & 0.023 &  & 0.940 & 0.940 & 0.940 \\ 
  ~~~~TBP & 0.009 & 0.009 & 0.009 &  & 0.023 & 0.023 & 0.023 &  & 0.922 & 0.922 & 0.922 \\ 
  ~~~~Weibull & 0.011 & 0.011 & 0.011 &  & 0.023 & 0.023 & 0.023 &  & 0.873 & 0.873 & 0.873 \\ 
   \multicolumn{3}{l}{~~\textit{Piecewise Effect Specification}} \\
  ~~~~Log-Normal & 0.003 & 0.000 & 0.001 &  & 0.025 & 0.027 & 0.046 &  & 0.933 & 0.940 & 0.957 \\ 
  ~~~~TBP & 0.000 & 0.015 & 0.037 &  & 0.025 & 0.026 & 0.046 &  & 0.950 & 0.905 & 0.864 \\ 
  ~~~~Weibull & -0.018 & 0.005 & 0.076 &  & 0.027 & 0.025 & 0.049 &  & 0.850 & 0.913 & 0.520 \\ 
   \multicolumn{3}{l}{~~\textit{Spline Effect Specification}} \\
  ~~~~Log-Normal & 0.004 & 0.001 & -0.003 &  & 0.026 & 0.027 & 0.044 &  & 0.927 & 0.933 & 0.970 \\ 
  ~~~~TBP & -0.005 & 0.011 & 0.031 &  & 0.026 & 0.026 & 0.043 &  & 0.923 & 0.887 & 0.887 \\ 
  ~~~~Weibull & -0.025 & 0.002 & 0.062 &  & 0.028 & 0.026 & 0.044 &  & 0.823 & 0.917 & 0.653 \\ 
  \multicolumn{3}{l}{\textbf{Piecewise True Effect}} \\
   \multicolumn{3}{l}{~~\textit{Constant Effect Specification}} \\
  ~~~~Log-Normal & 0.049 & -0.023 & -0.220 &  & 0.026 & 0.026 & 0.026 &  & 0.517 & 0.810 & 0.000 \\ 
  ~~~~TBP & 0.058 & -0.015 & -0.211 &  & 0.026 & 0.026 & 0.026 &  & 0.386 & 0.878 & 0.000 \\ 
  ~~~~Weibull & 0.078 & 0.005 & -0.191 &  & 0.027 & 0.027 & 0.027 &  & 0.187 & 0.897 & 0.000 \\ 
   \multicolumn{3}{l}{~~\textit{Piecewise Effect Specification}} \\
  ~~~~Log-Normal & 0.003 & 0.003 & 0.019 &  & 0.025 & 0.036 & 0.116 &  & 0.933 & 0.947 & 0.940 \\ 
  ~~~~TBP & -0.001 & 0.011 & 0.058 &  & 0.025 & 0.035 & 0.118 &  & 0.930 & 0.913 & 0.913 \\ 
  ~~~~Weibull & -0.018 & 0.006 & 0.114 &  & 0.027 & 0.034 & 0.123 &  & 0.847 & 0.920 & 0.757 \\ 
   \multicolumn{3}{l}{~~\textit{Spline Effect Specification}} \\
  ~~~~Log-Normal & 0.005 & -0.003 & 0.060 &  & 0.027 & 0.035 & 0.110 &  & 0.930 & 0.940 & 0.897 \\ 
  ~~~~TBP & -0.004 & 0.007 & 0.101 &  & 0.026 & 0.035 & 0.112 &  & 0.923 & 0.933 & 0.793 \\   
  ~~~~Weibull & -0.024 & -0.002 & 0.155 &  & 0.029 & 0.033 & 0.118 &  & 0.820 & 0.907 & 0.600 \\ 
   \hline
\end{tabular}}
\end{table}

\newpage

\section{Derivation of $V^{-1}(t\mid \bX)$ under Piecewise Linearity} \label{sec:vinvderiv}

Under piecewise linear specification of $V(t\mid \bX)$, define $J + 2$ knots $0=\tau_{0} < \tau_{1}<\dots<\tau_{J}<\tau_{J+1}=\infty$, with piecewise linear basis functions defined
$B_j(t\mid\btau) =t^{-1} (\min\{t,\tau_{j+1}\} - \tau_{j})_{+}$
where $(z)_{+} = \min\{0,z\}$. Assuming a flexible effect for $X_1$, the resulting specification becomes
\begin{align}
V(t\mid\bX) = t\times \exp\left(-\bX\trans\bbeta\right)\left[\sum_{j=1}^{J}\exp\left(-X_1\alpha_j\right) B_{j}(t\mid\btau) \right].
\end{align}
The inverse function $V^{-1}$ can be derived by inspection, noting that the inverse of an increasing piecewise linear function is also an increasing piecewise linear function, with changepoints shifted according to the values of $\bX$, $\bbeta$, and $\balpha$. Specifically, define $\btau^*$ such that $\tau_0^* = \tau_0 = 0$, $\tau_1^* =\tau_{1}\times \exp(-\bX\trans\bbeta)$, and for $j>1$,
\begin{align}
\tau^*_{j} & = \tau^*_1 + \sum_{l = 2}^j\exp(-\bX\trans\bbeta - X_1\alpha_{l-1})(\tau_{l} - \tau_{l - 1}).
\end{align}
The lines on each interval of $V^{-1}$ have the inverse slope of the line in the corresponding interval of $V$, so the final inverse function is succinctly written
\begin{align}
V^{-1}(t\mid\bX) = t\times \exp\left(\bX\trans\bbeta\right)\left[\sum_{j=1}^{J}\exp\left(X_1\alpha_j\right) B_{j}(t\mid\btau^*) \right].
\end{align}

\section{Derivation of Acceleration Factor for a Binary Time-Varying Covariate} \label{sec:tvcovderiv}

Let $X_1(t)$ be a binary-valued step function, such as an indicator for whether a non-terminal event has occurred by time $t$. Formally, define $X_1(t) = \bbI(t > t_{X})$, where $t_{X}$ is the time at which $X_1$ changes. Consider a single additional covariate time-invariant covariate $X_2$.

Notating $t_{X}^* = t_{X}\exp(-X_2\beta_2)$, the inverse of the covariate process $V$ defined in Equation (5.22)
is derived following Appendix~\ref{sec:vinvderiv} as
\begin{equation} \label{eq:tvVinvproc}
\begin{aligned}
V^{-1}(t\mid\bX(t)) & =  \exp\left(X_2\beta_2\right) \left[ \min(t,t_{X}^*) + (t - t_{X}^*)_{+} \exp\left(\beta_1\right) \right].
\end{aligned}
\end{equation}
The resulting acceleration factor at quantile $p$ between a person with $X_2=x_2$ who experiences the non-terminal event at time $t_X$, and a person with $X_2=x'_2$ who experiences the non-terminal event at time $t'_X$, is
\begin{equation} \label{eq:tvVinvproc}
\begin{aligned}
\xi(p\mid t_{X},t_{X}',x_2,x'_2,S_0) & =  e^{(x_2-x'_2)\beta_2}\frac{\min(S^{-1}_0(p),t_{X}e^{-x_2\beta_2}) + e^{\beta_1}(S^{-1}_0(p) - t_{X}e^{-x_2\beta_2})_{+}}{\min(S^{-1}_0(p),t_{X}'e^{-x'_2\beta_2}) + e^{\beta_1}(S^{-1}_0(p) - t_{X}'e^{-x'_2\beta_2})_{+}}.
\end{aligned}
\end{equation}
Finally, note that when a general flexible effect for $X_1(t)$ is specified, in general no closed form exists, but acceleration factors can still be computed numerically.


 
\newpage

\section{Bayes Factor for Comparison with Standard AFT} \label{sec:bayesfactor}

It may be of interest to compare AFT models with flexible covariate effects to standard AFT models, i.e., evaluating $H_0:\balpha=\bzero$ versus $H_1:\balpha\neq\bzero$, which is presented in the main paper through the use of the ELPD model metric. An alternative approach would be to use Bayes factors \citep{zhang2019bayes}, which we summarize here. Let $\bphi=(\mu,\sigma,\bw,\theta)\trans$ be the vector of parameters corresponding with the baseline survival distribution, and let $\Dsc$ denote the collected data, yielding likelihood function $\Lsc$ and posterior $p(\balpha,\bbeta,\bphi\mid\Dsc)$ derived from prior $p_0(\bbeta,\balpha,\bphi)$. The corresponding Bayes factor for $H_0$ relative to $H_1$ under these priors is
\[
\text{BF}=\frac{\iint \Lsc(\balpha=\bzero,\bbeta,\bphi)p_0(\bbeta,\bphi) d\bphi d\bbeta}{\iiint \Lsc(\balpha,\bbeta,\bphi)p_0(\balpha,\bbeta,\bphi) d\bphi d\bbeta d\balpha}
\]
where $p_0(\bbeta,\bphi) = \int p_0(\balpha,\bbeta,\bphi)d\balpha$ is the marginal prior on $\bbeta$ and $\bphi$. Larger values reflect evidence for $H_0$ over $H_1$. If the prior is specified such that \citep{verdinelli1995computing}
\begin{align}\label{ass:bayesfactor} 
p_0(\bbeta,\bphi) = p_0(\bbeta,\bphi\mid\balpha=\bzero),
\end{align}
then 
the Bayes factor can be expressed as a `Savage-Dickey' density ratio of the marginal posterior density of $\balpha$ and the marginal prior density of $\balpha$, each evaluated at $\bzero$:
\begin{align}\label{eq:savagedickey}
\text{BF} = \frac{\iint p(\balpha=\bzero,\bbeta,\bphi\mid\Dsc)d\bphi d\bbeta}{\int\int p_0(\balpha=\bzero,\bbeta,\bphi) d\bphi d\bbeta} = \frac{p(\balpha=\bzero\mid\Dsc)}{p_0(\balpha=\bzero)}.
\end{align}

The use of Bayes Factors requires proper priors on all parameters. Though in the main text we used improper `flat' priors on $\balpha$ and $\bbeta$, we might instead specify a multivariate normal prior 
\begin{align}\label{eq:mvnprior}
\begin{bmatrix} \bbeta \\ \balpha \end{bmatrix} \sim \MVN\left( 
\mathbf{m}_{0}=\begin{bmatrix} \mathbf{m}_{\bbeta,0} \\ \mathbf{m}_{\balpha,0} \end{bmatrix},  
\bV_{0}=\begin{bmatrix} \bV_{\bbeta\bbeta,0} &\bV_{\bbeta\balpha,0} \\ \bV_{\balpha\bbeta,0} & \bV_{\balpha\balpha,0} \end{bmatrix}
\right),
\end{align}
denoted $p_0(\bbeta,\balpha;\mathbf{m}_{0},\bV_{0})$. Note that by properties of multivariate normals, marginally \eqref{eq:mvnprior} implies $\bbeta \sim \MVN(\mathbf{m}_{\bbeta,0},\bV_{\bbeta\bbeta,0})$ and $\balpha \sim \MVN(\mathbf{m}_{\balpha,0},\bV_{\balpha\balpha,0})$, and conditional prior 
\[\bbeta \mid \balpha=\bzero \sim \MVN\left[ \left(\mathbf{m}_{\bbeta,0} - \bV_{\bbeta\balpha,0}\bV_{\balpha\balpha,0}^{-1}\mathbf{m}_{\balpha,0}\right), \left(\bV_{\bbeta\beta,0} - \bV_{\bbeta\balpha,0}\bV_{\balpha\balpha,0}^{-1}\bV_{\balpha\bbeta,0}\right) \right].
\]
Assuming the baseline parameters $\bphi$ have some prior $p_0(\bphi)$ which is independent of the priors for $\bbeta,\balpha$, then under a prior of the form \eqref{eq:mvnprior}, \eqref{ass:bayesfactor} holds if either $\mathbf{m}_{\balpha,0}=\bzero$, or $\bV_{\bbeta\balpha,0}$ and $\bV_{\balpha\bbeta,0}$ are zero matrices (i.e., priors on $\bbeta$ and $\balpha$ are independent).   \eqref{ass:bayesfactor} further extends to any proper independent priors on $\bbeta$ and $\balpha$, such as independent t-distribution or shrinkage priors.

Finally, \eqref{eq:savagedickey} can be estimated by approximating the marginal posterior density $p(\balpha \mid\Dsc)$ as the density of a multivariate normal distribution with mean and covariance matrix equal to the estimated posterior mean and covariance matrix of $\balpha$ \citep{zhang2019bayes}, divided by the marginal prior, each evaluated at $\balpha=\bzero$.

\newpage

\section{Transformed Bernstein Polynomial Prior} \label{sec:tbp_detail}

To illustrate the flexibility of the transformed Bernstein polynomial prior, Figure~\ref{fig:tbp_basis} shows the basis functions when $K=5$, \[G(p\mid k, K - k + 1) =  \frac{\Gamma(K+1)}{\Gamma(k)\Gamma(K-k+1)} p^{k-1}(1-p)^{K-k}.\] Moreover, Figure~\ref{fig:tbp_basis_ex} shows a sample of different shapes that the resulting baseline survivor function $S_0(t \mid \bphi, \bw) = \sum_{k = 1}^{K} w_k G(S^{*}_0(t \mid \bphi)\mid k, K - k + 1)$ can take, for selected weight vectors $\bw$ and setting $S^{*}_0(t) = \exp(-t)$.

\begin{figure}[H]
\begin{center}
\includegraphics[width=5.5in]{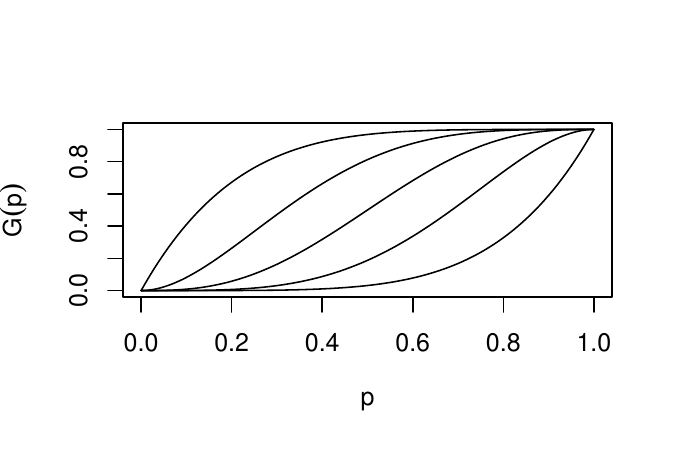}
\end{center}	\vspace{-0.6in}
\caption{Basis functions $G$ for $j = 1,\dots,5$. \label{fig:tbp_basis}}
\end{figure}

\begin{figure}[H]
\begin{center}
	\minipage{0.58\textwidth}%
	    \includegraphics[width=\linewidth]{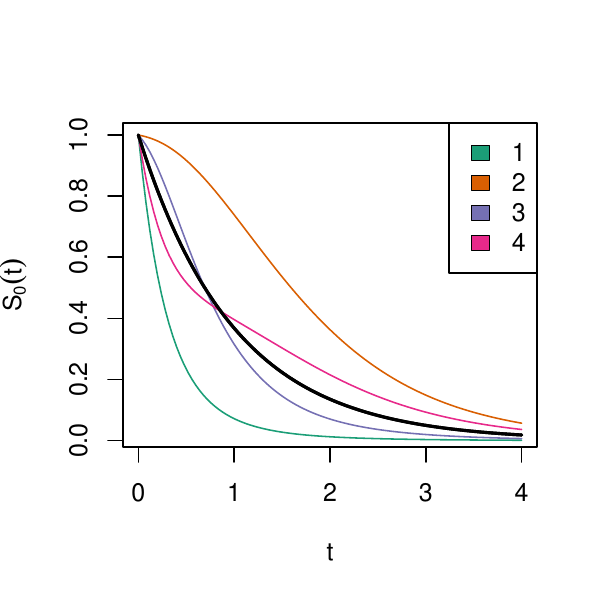}
	\endminipage\hfill
	\minipage{0.38\textwidth}
            \begin{tabular}{rrrrr}
              \hline
             & 1 & 2 & 3 & 4 \\ 
              \hline
            $w_1$ & 0.01 & 0.64 & 0.07 & 0.41 \\ 
              $w_2$ & 0.03 & 0.23 & 0.18 & 0.02 \\ 
              $w_3$ & 0.09 & 0.09 & 0.50 & 0.01 \\ 
              $w_4$ & 0.23 & 0.03 & 0.18 & 0.15 \\ 
              $w_5$ & 0.64 & 0.01 & 0.07 & 0.41 \\ 
               \hline
            \end{tabular}
	\endminipage
	\vspace{-0.2in}
\caption{Sample survivor functions corresponding to varying transformed Bernstein polynomial prior weight vectors. Bold black line shows centering distribution $S_0^*(t) = \exp(-t)$, equivalent to when weights are equal. \label{fig:tbp_basis_ex}}
\end{center}
\end{figure}

\end{document}